%% file: Cattaneo-Jansson-Ma_2018_ManyCovs.tex
\documentclass[11pt]{article} 
\usepackage{amsfonts,amssymb,amsmath,amsthm,dsfont,graphicx,hyperref,mathrsfs,euscript}
\usepackage[top=1in,bottom=1in,left=1in,right=1in]{geometry} 
\usepackage{caption}
\usepackage{xcolor}
\definecolor{umblue}{RGB}{0,85,148}
\usepackage{verbatim}
\usepackage{booktabs}
\usepackage{multirow}
\usepackage{framed}
\usepackage{forloop}
\usepackage{lscape}
\usepackage[onehalfspacing]{setspace}
\usepackage{tocloft} 
\usepackage{natbib}
\usepackage{ifthen}
\usepackage{subfig}

\hypersetup{pdfborder = {0 0 0},colorlinks=true,linkcolor=blue,urlcolor=blue,citecolor=blue}
\usepackage{listings}
\definecolor{codegreen}{rgb}{0,0.6,0}
\definecolor{codegray}{rgb}{0.5,0.5,0.5}
\definecolor{codepurple}{rgb}{0.58,0,0.82}
\definecolor{backcolour}{rgb}{0.95,0.95,0.92}

\lstdefinestyle{mystyle}{
	language=R,
    backgroundcolor=\color{backcolour},   
    commentstyle=\color{codegreen},
    numberstyle=\tiny\color{codegray},
    basicstyle=\footnotesize\ttfamily,
    breakatwhitespace=false,         
    breaklines=true,                 
    captionpos=b,                    
    keepspaces=true,                 
    numbers=left,                    
    numbersep=5pt,                  
    showspaces=false,                
    showstringspaces=false,
    showtabs=false,                  
    tabsize=2
}
 
\newtheoremstyle{exampstyle}
  {} 
  {} 
  {} 
  {} 
  {\bfseries\color{blue}} 
  {.} 
  {.5em} 
  {} 
\theoremstyle{exampstyle}\newtheorem{thm}{Theorem}
\theoremstyle{exampstyle}\newtheorem*{thm*}{Theorem}
\theoremstyle{exampstyle}     
\theoremstyle{exampstyle}     
\theoremstyle{exampstyle}  
\theoremstyle{exampstyle}\newtheorem{coro}{Corollary}        
\theoremstyle{exampstyle}

\theoremstyle{exampstyle}\newtheorem{Assumption}{Assumption}
\theoremstyle{definition}

\theoremstyle{exampstyle}
\newtheorem{remark_tmp}{Remark}
\newenvironment{remark}
	{ \begin{remark_tmp} 	}
	{ 
		\medskip\hfill{\LARGE$\lrcorner$}
		\end{remark_tmp} 
	}

\allowdisplaybreaks[1]

\newtheorem*{example_tmp}{Example}

\allowdisplaybreaks[1]

\newcommand{\Expectation}{\mathbb{E}}
\newcommand{\Cov}{\mathbb{C}\mathrm{ov}}
\newcommand{\Var}{\mathbb{V}}

\newcommand{\Prob}{\mathbb{P}}
\newcommand{\Indicator}{\mathds{1}}
\newcommand{\op}{o_{\Prob}}
\newcommand{\Op}{O_{\Prob}}

\newcommand{\Trans}{\mathsf{T}}

\renewcommand{\epsilon}{\varepsilon}

\newcommand{\bM}{\mathbf{M}}

\newcommand{\bR}{\mathbf{R}}

\newcommand{\bW}{\mathbf{W}}
\newcommand{\bX}{\mathbf{X}}

\newcommand{\bZ}{\mathbf{Z}}

\newcommand{\bb}{\mathbf{b}}

\newcommand{\bg}{\mathbf{g}}
\newcommand{\bm}{\mathbf{m}}

\newcommand{\bw}{\mathbf{w}}
\newcommand{\bx}{\mathbf{x}}
\newcommand{\by}{\mathbf{y}}
\newcommand{\bz}{\mathbf{z}}

\newcommand{\bbeta}{\boldsymbol{\beta}}
\newcommand{\bdelta}{\boldsymbol{\delta}}

\newcommand{\blambda}{\boldsymbol{\lambda}}
\newcommand{\bmu}{\boldsymbol{\mu}}
\newcommand{\bgamma}{\boldsymbol{\gamma}}

\newcommand{\bphi}{\boldsymbol{\phi}}
\newcommand{\btheta}{\boldsymbol{\theta}}
\newcommand{\bzeta}{\boldsymbol{\zeta}}

\newcommand{\bGamma}{\boldsymbol{\Gamma}}
\newcommand{\bSigma}{\boldsymbol{\Sigma}}
\newcommand{\bPsi}{\boldsymbol{\Psi}}

\newcommand{\bPi}{\boldsymbol{\Pi}}

\newcommand{\MTE}{\tau_{\mathtt{MTE}}}

\DeclareMathOperator*{\argmin}{argmin} 

\begin{document}

\title{\vspace{-0.5in} Two-Step Estimation and Inference\\ with Possibly Many Included Covariates\thanks{
This paper encompasses and supersedes our previous paper titled ``Marginal Treatment Effects with Many Instruments'', presented at the 2016 NBER summer meetings. We specially thank Pat Kline for posing a question that this paper answers, and Josh Angrist, Guido Imbens and Ed Vytlacil for very useful comments on an early version of this paper. We also thank the Editor, Aureo de Paula, three anonymous reviewers, Lutz Kilian, Whitney Newey and Chris Taber for very useful comments. The first author gratefully acknowledges financial support from the National Science Foundation (SES 1459931). The second author gratefully acknowledges financial support from the National Science Foundation (SES 1459967) and the research support of CREATES (funded by the Danish National Research Foundation under grant no. DNRF78). Disclaimer: This research was conducted with restricted access to Bureau of Labor Statistics (BLS) data. The views expressed here do not necessarily reflect the views of the BLS.
}\bigskip}
\author{
Matias D. Cattaneo\thanks{Department of Economics and Department of Statistics, University of Michigan.} \and 
Michael Jansson\thanks{Department of Economics, UC Berkeley, and \textit{CREATES}.} \and 
Xinwei Ma\thanks{Department of Economics, University of Michigan.}
}\maketitle

\vspace{0.0in}\begin{abstract}
We study the implications of including many covariates in a first-step estimate entering a two-step estimation procedure. We find that a first order bias emerges when the number of \textit{included} covariates is ``large'' relative to the square-root of sample size, rendering standard inference procedures invalid. We show that the jackknife is able to estimate this ``many covariates'' bias consistently, thereby delivering a new automatic bias-corrected two-step point estimator. The jackknife also consistently estimates the standard error of the original two-step point estimator. For inference, we develop a valid post-bias-correction bootstrap approximation that accounts for the additional variability introduced by the jackknife bias-correction. We find that the jackknife bias-corrected point estimator and the bootstrap post-bias-correction inference perform excellent in simulations, offering important improvements over conventional two-step point estimators and inference procedures, which are not robust to including many covariates. We apply our results to an array of distinct treatment effect, policy evaluation, and other applied microeconomics settings. In particular, we discuss production function and marginal treatment effect estimation in detail.
\end{abstract}

\thispagestyle{empty}
\clearpage

\doublespacing
\setcounter{page}{1}
\pagestyle{plain}

\pagestyle{plain}

\section{Introduction}\label{section: introduction}

Two-step estimators are very important and widely used in empirical work in Economics and other disciplines. This approach involves two estimation steps: first an unknown quantity is estimated, and then this estimate is plugged in a moment condition to form the second and final point estimator of interest. For example, inverse probability weighting (IPW) and generated regressors methods fit naturally into this framework, both used routinely in treatment effect and policy evaluation settings. In practice, researchers often include many covariates in the first-step estimation procedure in an attempt to flexibly control for as many confounders as possible, even after model selection or model shrinking has been used to select out some of all available covariates. Conventional (post-model selection) estimation and inference results in this context, however, assume that the number of covariates included in the estimation is ``small'' relative to the sample size, and hence the effect of overfitting in the first estimation step is ignored in current practice.

We show that two-step estimators can be severely biased when too many covariates are included in a linear-in-parameters first-step, a fact that leads to invalid inference procedures even in large samples. This crucial, but often overlooked fact implies that many empirical conclusions will be incorrect whenever many covariates are used. For example, we find from a very simple simulation setup with a first step estimated with $80$ i.i.d. variables, sample size of $2,000$, and even no misspecification bias, that a conventional $95\%$ confidence interval covers the true parameter with probability $76\%$ due to the presence of the many covariates bias we highlight in this paper (Table \ref{table: MTE-2-BT-JK} below).\footnote{Including $80$ regressors is quite common in empirical work: e.g., settings with $50$ residential dummy indicators, a few covariates entering linearly and quadratically, and perhaps some interactions among these variables. The Supplemental Appendix discusses several examples employing two-step estimators with possibly many covariates.} This striking result is not specific to our simulation setting, as our general results apply broadly to many other treatment effect, policy evaluation, and applied microeconomics settings: IPW estimation under unconfoundedness, semiparametric difference-in-differences, local average response function estimation, marginal treatment effects, control function methods, and production function estimation, just to mention a few other popular examples.

We illustrate the usefulness of our results by considering several applications in applied microeconomics. In particular, we discuss in detail production function \citep{Olley-Pakes-1996_ECMA} and marginal treatment effect \citep{Heckman-Vytlacil_2005_ECMA} estimation when possibly many  covariates/instruments are present. The latter application offers new estimation and inference results in instrumental variable (IV) settings allowing for treatment effect heterogeneity and many covariates/instruments.

The presence of the generic many covariates bias we highlight implies that developing more robust procedures accounting for possibly many covariates entering the first step estimation is highly desirable. Such robust methods would give more credible empirical results, thereby providing more plausible testing of substantive hypotheses as well as more reliable policy prescriptions. With this goal in mind, we show that jackknife bias-correction is able to remove the many covariates bias we uncover in a fully automatic way. Under mild conditions on the design matrix, we prove consistency of the jackknife bias and variance estimators, even when many covariates are included in the first-step estimation. Indeed, our simulations in the context of MTE estimation show that jackknife bias-correction is quite effective in removing the many covariates bias, exhibiting roughly a $50\%$ bias reduction (Table \ref{table: MTE-2-BT-JK} below). We also show that the mean squared error of the jackknife bias-corrected estimator is substantially reduced whenever many covariates are included. More generally, our results give a new, fully automatic, jackknife bias-corrected two-step estimator with demonstrably superior properties to use in applications.

For inference, while the jackknife bias correction and variance estimation deliver a valid Gaussian distributional approximation in large samples, we find in our simulations that the associated inference procedures do not perform as well in small samples. As discussed in \citet{Calonico-Cattaneo-Farrell_2018_JASA} in the context of kernel-based nonparametric inference, a crucial underlying issue is that bias correction introduces additional variability not accounted for in samples of moderate size (we confirm this finding in our simulations). Therefore, to develop better inference procedures in finite samples, we also establish validity of a bootstrap method applied to the jackknife-based bias-corrected Studentized statistic, which can be used to construct valid confidence intervals and conduct valid hypothesis tests in a fully automatic way. This procedure is a hybrid of the wild bootstrap (first-step estimation) and the multiplier bootstrap (second-step estimation), which is fast and easy to implement in practice because it avoids recomputing the relatively high-dimensional portion of the first estimation step. Under generic regularity conditions, we show that this bootstrap procedure successfully approximates the finite sample distribution of the bias-corrected jackknife-based Studentized statistic, a result that is also borne out in our simulation study.

Put together, our results not only highlight the important negative implications of overfitting the first-step estimate in generic two-step estimation problems, which leads to a first order many covariates bias in the distributional approximation, but also provide fully automatic resampling methods to construct more robust estimators and inference procedures. Furthermore, because our results remain asymptotically valid when only a few covariates are used, they provide strict asymptotic improvement over conventional methods currently used in practice. All our results are fully automatic and do not require additional knowledge about the data generating process, which implies that they can be easily implemented in empirical work using straightforward resampling methods on any computing platform.

In the remainder of this introduction section we discuss some of the many literatures this paper is connected to. Then, the rest of the paper unfolds as follows. Section \ref{section: Overview} introduces the setup and gives an overview of our results. Section \ref{section: asymptotic} gives details on the main properties of the two-step estimator, in particular characterizing the non-vanishing bias due to many covariates entering the first-step estimate. Section \ref{section: jackknife} establishes validity of the jackknife bias and variance estimator, and therefore presents our proposed bias-corrected two-step estimator, while Section \ref{section: bootstrap} establishes valid distributional approximations for the jackknife-based bias-corrected Studentized statistic using a carefully modified bootstrap method. Section \ref{section: examples} applies our main results to two examples: production function \citep{Olley-Pakes-1996_ECMA} and marginal treatment effect \citep{Heckman-Vytlacil_2005_ECMA} estimation, while six other treatment effect, program evaluation and applied microeconomics examples are discussed in the Supplemental Appendix (Section SA-5) to conserve space. Section \ref{section: numerical} summarizes the main results from an extensive Monte Carlo experiment and an empirical illustration building on the work of \citet{Carneiro-Heckman-Vytlacil_2011_AER}. Finally, Section \ref{section: conclusion} concludes. The Supplemental Appendix also contains methodological and technical details, includes the theoretical proofs of our main theorems as well as other related results, and reports additional numerical evidence.

\subsection{Related Literature}

Our work is related to several interconnected literatures in econometrics and statistics.

\textit{Two-step Semiparametrics}.
From a classical semiparametric perspective, when the many covariates included in the first-step are taken as basis expansions of some underlying fixed dimension regressor, our final estimator becomes a two-step semiparametric estimator with a nonparametric series-based preliminary estimate. Conventional large sample approximations in this case are well known \citep[e.g.,][and references therein]{Newey-McFadden_1994_Handbook,Chen_2007_Handbook,Ichimura-Todd_2007_Handbook}. From this perspective, our paper contributes not only to this classical semiparametric literature, but also to the more recent work in the area, which has developed distributional approximations that are more robust to tuning parameter choices and underlying assumptions (e.g., smoothness). In particular, first, \citet{Cattaneo-Crump-Jansson_2013_JASA} and \citet{Cattaneo-Jansson_2018_ECMA} develop approximations for two-step non-linear kernel-based semiparametric estimators when possibly a ``small'' bandwidth is used, which leads to a first-order bias due to undersmoothing the preliminary kernel-based nonparametric estimate, and show that inference based on the nonparametric bootstrap automatically accounts for the small bandwidth bias explicitly, thereby offering more robust inference procedures in that context.\footnote{A certain class of \textit{linear} semiparametric estimators has a very different behavior when undersmoothing the first step nonparametric estimator; see \citet{Cattaneo-Crump-Jansson_2010_JASA,Cattaneo-Crump-Jansson_2014a_ET,Cattaneo-Crump-Jansson_2014b_ET} and \citet{Cattaneo-Jansson-Newey_2018_ET} for discussion and references. In particular, their results show that undersmoothing leads to an additional variance contribution (due to the underlying linearity of the model), while in the present paper we find a bias contribution instead (due to the non-linearity of the models considered).} Second, \citet{Chernozhukov-Escanciano-Ichimura-Newey-Robins_2018_LocRobust} study the complementary issue of ``large'' bandwidth or ``small'' number of series terms, and develop more robust inference procedures in that case. Their approach is to modify the estimating equation so that the resulting new two-step estimator is less sensitive to oversmoothing (i.e., underfitting) the first-step nonparametric estimator. Our paper complements this recent by offering new inference procedures with demonstrably more robust properties to undersmoothing (i.e., overfitting) a first step series-based estimator, results that are not currently available in the semiparametrics literature. See Section \ref{section: asymptotic} for more details.

\textit{High-dimensional Models}.
Our results go beyond semiparametrics because we do not assume (but allow for) the first-step estimate to be a nonparametric series-based estimator. In fact, we do not rely on any specific structure of the covariates in the first step, nor do we rely on asymptotic linear representations. Thus, our results also contribute to the literature on high-dimensional models in statistics and econometrics \citep[e.g.,][and references therein]{Mammen_1989_AoS,Mammen_1993_AoS,ElKaroui-Bean-Bickel-Lim-Yu_2013_PNAS,Cattaneo-Jansson-Newey_2018_JASA,Li-Muller_2017_wp} by developing generic distributional approximations for two-step estimators where the first-step estimator is possibly high-dimensional. See also \cite{Fan-Lv-Qi_2011_ARE} for a survey and discussion on high-dimensional and ultra-high-dimensional models.\footnote{We call models high-dimensional when the number of available covariates is at most a fraction of the sample size and ultra-high-dimensional when the number of available covariates is larger than the sample size.} A key distinction here is that the class of estimators we consider is defined through a moment condition that is non-linear in the first step estimate (e.g., propensity score, generated regressor, etc.). Previous work on high-dimensional models has focused exclusively on either linear least squares regression or one-step (possibly non-linear) least squares regression. In contrast, this paper covers a large class of two-step non-linear procedures, going well beyond least squares regression for the second step estimation procedure. Most interestingly, our results show formally that when many covariates are included in a first-step estimation the resulting two-step estimator exhibits a bias of order $k/\sqrt{n}$ in the distributional approximation, where $k$ denotes the number of included covariates and $n$ denotes the sample size. This finding contrasts sharply with previous results for high-dimensional linear regression models with many covariates, where it has been found that including many covariates leads to a variance contribution (not a bias contribution as we find herein) in the distributional approximation, which is of order $k/n$ (not $k/\sqrt{n}$ as we find herein). By implication, the many covariates bias we uncover in this paper will have a first-order effect on inference when fewer covariates are included relative to the case of high-dimensional linear regression models.

\textit{Ultra-high-dimensional Models and Covariate Selection}.
Our results also have implications for the recent and rapidly growing literature on inference after covariate/model selection in ultra-high-dimensional settings under sparsity conditions \citep[e.g.,][and references therein]{Belloni-Chernozhukov-Hansen_2014_ReStud,Farrell_2015_JoE,Belloni-Chernozhukov-FernandezVal-Hansen_2017_ECMA}. In this literature, the total number of available covariates/instruments is allowed to be much larger than the sample size, but the final number of \textit{included} covariates/instruments is much smaller than the sample size, as most available covariates are selected out by some penalization or model selection method (e.g., LASSO) employing some form of a sparsity assumption. This implies that the number of included covariates/instruments effectively used for estimation and inference ($k$ in our notation) is much smaller than the sample size, as the underlying distribution theory in that literature requires $k/\sqrt{n}=o(1)$. Therefore, because $k/\sqrt{n}=O(1)$ is the only restriction assumed in this paper, our results shed new light on situations where the number of selected or included covariates, possibly after model selection, is not ``small'' relative to the sample size. We formally show that valid inference post-model selection requires that a relatively small number of covariates enter the final specification, since otherwise a first order bias will be present in the distributional approximations commonly employed in practice, thereby invalidating the associated inference procedures. Our results do not employ any sparsity assumption and allow for any kind of regressors, including many fixed effects, provided the first-step estimate can be computed.

\textit{Large-($N,T$) Panel Data}.
Our findings are also qualitatively connected to the literature on non-linear panel models with fixed effects \citep[and references therein]{FernandezVal-Weidner_2018_ARE} in at least two ways. First, in that context a first-order bias arises when the number of time periods ($T$) is proportional to the number of entities ($N$), just like we uncover a first-order bias when $k\propto\sqrt{n}$, and in both cases this bias can be heuristically attributed to an incidental parameters/overfitting problem. Second, in that literature jackknife bias correction was shown to be able to remove the large-($N,T$) bias, just like we establish a similar result in this paper for a class of two-step estimators with high-dimensional first-step. Beyond these two superficial connections, however, our findings are both technically and conceptually quite different from the results already available in the large-($N,T$) non-linear panel fixed effects literature.

\textit{Applications}.
From a practical perspective, our results offer new inference results for many popular estimators in program evaluation and treatment effect settings \citep[e.g.,][and references therein]{Abadie-Cattaneo_2018_ARE}, as well as other areas in empirical microeconomics \citep[e.g.,][and references therein]{Ackerberg-Benkard-Berry-Pakes_2007_Handbook}. Section \ref{section: Olley-Pakes} discusses production function estimation, which provides new econometric methodology in the context of an IO application, while Section \ref{section: MTE} considers marginal treatment effect estimation, where we develop new estimation and inference methods in the presence of many covariates/instruments and heterogeneous IV treatment effects. Furthermore, because our results apply to non-linear settings in general, we cover many other settings of interest: (i) IPW under unconfoundedness \cite[e.g.,][and references therein]{Cattaneo_2010_JoE}, (ii) Semiparametric Difference-in-Differences \citep{Abadie_2005_RES}, (iii) Local Average Response Function \citep{Abadie_2003_JoE}, (iv) Two-Stage Least Squares and Conditional Moment Restrictions \citep[e.g.,][for a textbook review]{Wooldridge_2010_Book}, and (v) Control Function Methods \citep[e.g.,][and references therein]{Wooldridge_2015_JHR}. All these other examples are analyzed in Section SA-5 of the Supplemental Appendix.

\section{Setup and Overview of Results}\label{section: Overview}

We consider a two-step GMM setting where $\bw_i=(\by_i^\Trans,r_i,\bz_i^\Trans)^\Trans$, $i=1,2,\dots,n$, denotes an observed random sample, and the finite dimensional parameter of interest $\btheta_0$ solves uniquely the (possibly over-identified) vector-valued moment condition $\Expectation[\bm(\bw_i,\mu_i,\btheta_0)]=\mathbf{0}$ with $\mu_i=\Expectation[r_i|\bz_i]$. Thus, we specialize the general two-step GMM approach in that we view the unknown scalar $\mu_i$ as a ``generated regressor'' depending on possibly many covariates $\bz_i\in\mathbb{R}^{k}$, which we take as the included variables entering the first-step specification. Our results extend immediately to vector-valued unknown $\bmu_i$, albeit with cumbersome notation, as shown in Section SA-4.1 of the Supplemental Appendix. See Section \ref{section: Olley-Pakes} for an application with a bivariate first-step.

Given a first-step estimate $\hat{\mu}_i$ of $\mu_i$, which we construct by projecting $r_i$ on the possibly high-dimensional covariate $\bz_i$ with least squares, as discussed further below, we study the two-step estimator:
\begin{align}
\label{eq: second step, estimation}
\hat{\btheta} = \arg\min_{\btheta\in\Theta}\left|\boldsymbol{\Omega}_n^{1/2}\sum_{i=1}^{n} \bm(\bw_i, \hat{\mu}_i,\btheta)\right|,
\end{align}
where $|\cdot|$ denotes the Euclidean norm, $\Theta\subseteq\mathbb{R}^{d_\theta}$ is the parameter space, and $\boldsymbol{\Omega}_n$ is a (possibly random) positive semi-definite conformable weighting matrix with positive definite probability limit $\boldsymbol{\Omega}_0$. Regularity conditions on the known moment function $\bm(\cdot)$ are given in the next section. 

When the dimension of the included variables $\bz_i$ is ``small'' relative to the sample size, $k=o(\sqrt{n})$, textbook large sample theory is valid, and hence estimation and inference can be conducted in the usual way \citep[e.g.,][]{Newey-McFadden_1994_Handbook}. However, when the dimension of the included covariates used to approximate the unknown component $\mu_i$ is ``large'' relative to the sample size, $k=O(\sqrt{n})$, standard distribution theory fails. To be more specific, under fairly general regularity conditions, we show in Section \ref{section: asymptotic} that:
\begin{align}\label{eq:Result-1}
  \mathscr{V}^{-1/2} (\hat{\btheta} - \btheta_0 - \mathscr{B})
  \ \rightsquigarrow\ \mathrm{Normal}(\mathbf{0},\ \mathbf{I}),
\end{align}
where $\rightsquigarrow$ denotes convergence in distribution, with limits always taken as $n\to\infty$ and $k=O(\sqrt{n})$, and $\mathscr{V}$ and $\mathscr{B}$ denoting, respectively, the approximate variance and bias of the estimator $\hat{\btheta}$. This result has a key distinctive feature relative to classical textbook results: a first-order bias $\mathscr{B}$ emerges whenever ``many'' covariates are included, that is, whenever $k$ is ``large'' relatively to $n$ in the sense that $k/\sqrt{n}\not\to 0$. A crucial practical implication of this finding is that conventional inference procedures that disregard the presence of the first-order bias will be incorrect even asymptotically, since $\mathscr{V}^{-1/2}\mathscr{B}=\Op(k/\sqrt{n})$ is non-negligible. For example, non-linear treatment effect, instrumental variables and control function estimators employing ``many'' included covariates in a first-step estimation will be biased, thereby giving over-rejection of the null hypothesis of interest. In Section \ref{section: numerical} we illustrate this problem using simulated data in the context of instrumental variable models with many instruments/covariates, where we find that typical hypothesis tests over-reject the null hypothesis four times as often as they should in practically relevant situations.

Putting aside the bias issue when many covariates are used in the first-step estimation, another important issue regarding (\ref{eq:Result-1}) is the characterization and estimation of the variance $\mathscr{V}$. Because the possibly high-dimensional covariates $\bz_i$ are not necessarily assumed to be a series expansion, or other type of convergent sequence of covariates, the variance $\mathscr{V}$ is harder to characterize and estimate. In fact, our distributional approximation leading to (\ref{eq:Result-1}) is based on a quadratic approximation of $\hat{\btheta}$, as opposed to the traditional linear approximation commonly encountered in the semiparametrics literature \citep{Newey_1994_ECMA,Chen_2007_Handbook,Hahn-Ridder_2013_ECMA}, thereby giving a more general characterization of the variability of $\hat{\btheta}$ with potentially better finite sample properties.

Nevertheless, our first main result (\ref{eq:Result-1}) suggests that valid inference in two-step GMM settings is possible even when many covariates are included in the first-step estimation, if consistent variance and bias estimators are available. Our second main result (in Section \ref{section: jackknife}) shows that the jackknife offers an easy-to-implement and automatic way to approximate both the variance and the bias:
\begin{align}\label{eq:Result-2}
\mathscr{T}\ \overset{\mathrm{def}}{=}\ 
\hat{\mathscr{V}}^{-1/2} (\hat{\btheta} - \btheta_0  - \hat{\mathscr{B}})\ 
\rightsquigarrow\ \mathrm{Normal}(\mathbf{0},\ \mathbf{I}),
\end{align}
\begin{align}\label{eq:Bias-Variance}
\hat{\mathscr{B}} = (n-1) (\hat{\btheta}^{(\cdot)}- \hat{\btheta}), \qquad
\hat{\mathscr{V}} = \frac{n-1}{n} \sum_{\ell=1}^n (\hat{\btheta}^{(\ell)} - \hat{\btheta}^{(\cdot)})
(\hat{\btheta}^{(\ell)} - \hat{\btheta}^{(\cdot)})^\Trans, \qquad
\hat{\btheta}^{(\cdot)} = \frac{1}{n}\sum_{\ell=1}^n\hat{\btheta}^{(\ell)},
\end{align}
where to construct $\hat{\btheta}^{(\ell)}$, the $\ell^{\text{th}}$ observation is deleted and then both steps are re-estimated using the remaining observations. Simulation evidence reported in Section \ref{section: numerical} confirms that the jackknife provides an automatic data-driven method able to approximate quite well both the bias and the variance of the estimator $\hat{\btheta}$, even when many covariates are included in the first-step estimation procedure. An important virtue of the jackknife is that it can be implemented very fast in special settings, which is particularly important in high-dimensional situations. Indeed, our first-step estimator will be constructed using least-squares, a method that is particularly amenable to jackknifing.

While result (\ref{eq:Result-2}) could be used for inference in large samples, a potential drawback is that the jackknife bias-correction introduces additional variability not accounted for in samples of moderate size. Therefore, to improve inference further in applications, we develop a new, specifically tailored bootstrap-based distributional approximation to the jackknife-based bias-corrected and Studentized statistic. Our method combines the wild bootstrap (first-step) and the multiplier bootstrap (second-step), while explicitly taking into account the effect of jackknifing under the multiplier bootstrap law (see Section \ref{section: bootstrap} for more details). To be more specific, our third and final main result is:
\begin{align}\label{eq:Result-3}
\sup_{t\in\mathbb{R}^{d_\theta}} \Big| \mathbb{P}[\mathscr{T}\leq t] - \mathbb{P}^\star[\mathscr{T}^\star\leq t] \Big| \to_\mathbb{P} 0, \qquad
\mathscr{T}^\star\ \overset{\mathrm{def}}{=}\ 
\hat{\mathscr{V}}^{\star^{-1/2}} (\hat{\btheta}^\star - \hat{\btheta} - \hat{\mathscr{B}}^\star),
\end{align}
where $\hat{\btheta}^\star$ is a bootstrap counterpart of $\hat{\btheta}$, $\hat{\mathscr{B}}^\star$ and $\hat{\mathscr{V}}^\star$ are properly weighted jackknife bias and variance estimators under the bootstrap distribution, respectively, and $\mathbb{P}^\star$ is the bootstrap probability law conditional on the data. Our bootstrap approach is fully automatic and captures explicitly the distributional effects of estimating the bias (and variance) using the jackknife, and hence delivers a better finite sample approximation. Simulation evidence reported in Section \ref{section: numerical} supports this result.

In sum, valid and more robust inference in two-step GMM settings with possibly many covariates entering the first-step estimate can be conducted by combining results (\ref{eq:Result-2}) and (\ref{eq:Result-3}). Specifically, our approach requires three simple and automatic stages: (i) constructing the two-step estimator ($\hat{\btheta}$), (ii) constructing the jackknife bias and variance estimators ($\hat{\mathscr{B}}$ and $\hat{\mathscr{V}}$), and finally (iii) conducting inference as usual but employing bootstrap quantiles obtained from (\ref{eq:Result-3}) instead of those from the normal approximation. In the remainder of this paper we formalize these results and illustrate them using simulated as well as real data.

\section{The Effect of Including Many Covariates\label{section: asymptotic}}

In this section we formalize the implications of overfitting the first-step estimate entering (\ref{eq: second step, estimation}), and show that under fairly general conditions the estimator $\hat{\btheta}$, and transformations thereof, exhibit a first-order bias whenever $k$ is ``large'', that is, whenever $k\propto\sqrt{n}$. The results in this section justify, in particular, the distributional approximation in (\ref{eq:Result-1}).

\subsection{Regularity Conditions}

A random variable is said to be in $\mathsf{BM}_\ell$ (bounded moments) if its $\ell^{\text{th}}$ moment is finite, and in $\mathsf{BCM}_\ell$ (bounded conditional moments) if its $\ell^{\text{th}}$ conditional on $\bz_i$ moment is bounded uniformly by a finite constant. In addition, define the transformation
\[\mathcal{H}_i^{\alpha,\delta}(\bm) = \sup_{(|\mu-\mu_i|+|\btheta-\btheta_0|)^\alpha\leq \delta} \frac{|\bm(\bw_i,\mu,\btheta) - \bm(\bw_i,\mu_i,\btheta_0)|}{(|\mu-\mu_i|+|\btheta-\btheta_0|)^\alpha}.\]
The following assumption collects some basic notation and regularity conditions.

\begin{Assumption}[Regularity Conditions]\label{Assumption 2: Smoothness and Moments} Let $0<\delta$, $\alpha$, $C<\infty$ be some fixed constants. \ \\
	\indent (i) $\bm$ is continuously differentiable in $\btheta$, and twice continuously differentiable in $\mu$ with derivatives denoted by $\dot{\bm}(\bw_i, \mu, \btheta_0) = \frac{\partial}{\partial \mu} \bm(\bw_i, \mu, \btheta_0)$ and $\ddot{\bm}(\bw_i, \mu, \btheta_0) = \frac{\partial^2}{\partial \mu^2} \bm(\bw_i, \mu, \btheta_0)$.
	\\
	\indent (ii) $\mathcal{H}_i^{\alpha,\delta}(\bm),\mathcal{H}_i^{\alpha,\delta}(\frac{\partial\bm}{\partial \btheta}),\mathcal{H}_i^{\alpha,\delta}(\frac{\partial\dot{\bm}}{\partial \btheta})\in\mathsf{BM}_1$.
	\\
	\indent (iii) ${\bm}_i$, $\dot{\bm}_i$, $\ddot{\bm}_i$, $\mathcal{H}_i^{\alpha,\delta}(\ddot{\bm})$, $\epsilon_i^2$, $|\dot{\bm}_i\epsilon_i|$, $|\ddot{\bm}_i|\epsilon_i^2$, $|\mathcal{H}_i^{\alpha,\delta}(\ddot{\bm})|\epsilon_i^2\in\mathsf{BCM}_2$, where ${\bm}_i=\bm(\bw_i,\mu_i,\btheta)$, $\dot{\bm}_i=\dot{\bm}(\bw_i,\mu_i,\btheta)$, $\ddot{\bm}_i=\ddot{\bm}(\bw_i,\mu_i,\btheta)$, and $\epsilon_i = r_i - \mu_i$.
	\\
	\indent (iv) $\bM_0=\Expectation\left[\frac{\partial}{\partial \btheta}\bm(\bw_i, \mu_i, \btheta_0)\right]$ has full (column) rank $d_\theta$.
\end{Assumption}

These conditions are standard in the literature. They require smoothness of $\bm(\bw,\mu,\btheta)$ with respect to both $\mu$ and $\btheta$, and boundedness of (conditional) moments of various orders. In future work we plan to extend our results to non-differentiable second-step estimating equations.

\subsection{First-Step Estimation}

We are interested in understanding the effects of including possibly many covariates $\bz_i$, that is, in cases where its dimension $k$ is possibly ``large'' relative to the sample size. For tractability and simplicity, we consider linear approximations to the unknown component:
\begin{align}
\label{eq: first step, population}
  \mu_i  = \Expectation[r_i|\bz_i] = \bz_i^\Trans\bbeta + \eta_i,\qquad
  \Expectation[\bz_i\eta_i]=\mathbf{0},
\end{align}
for a non-random vector $\bbeta$, where $\eta_i$ represents the error in the best linear approximation. This motivates the least-squares first-step estimate:
\begin{align}
\label{eq: first step, estimation}
\hat{\mu}_i &= \bz_i^\Trans\hat{\bbeta},\qquad \hat{\bbeta}\in\argmin_{\bbeta\in\mathbb{R}^k}\sum_{i=1}^{n}( r_i - \bz_i^\Trans\bbeta)^2,
\end{align}
which is quite common in empirical work. It is possible to allow for non-linear models, but such methods are harder to handle mathematically and usually do not perform well numerically when $\bz_i$ is of large dimension. Furthermore, a non-linear approach will be computationally more difficult, as we discuss in more detail below. As shown in the already lengthy Supplemental Appendix (Section SA-9), our proofs explicitly exploit the linear regression representation of $\hat{\mu}_i$ to scale down the already quite involved technical work. Nevertheless, we also conducted preliminary theoretical work to verify that the main results presented below carry over to non-linear least-squares estimators (e.g., logistic regression when $r_i$ is binary). 

Using the first-step estimate $\hat{\mu}_i$ in (\ref{eq: first step, estimation}), we investigate the implications of introducing possibly many covariates $\bz_i$, and thus our approximations allow for (but do not require that) $k$ being ``large'' relative to the sample size. Specifically, we show that when $k\propto\sqrt{n}$ conventional inference procedures become invalid due to a new bias term in the asymptotic approximations.

In some settings, the covariates $\bz_i$ can have approximation power beyond the first-step estimation, as it occurs for instance when these covariates are basis expansions. To allow for this possibility, we also define, for a non-random matrix $\bGamma$,
\begin{align}
\label{eq: other first step, population}
\Expectation[\dot{\bm}(\bw_i, \mu, \btheta_0)|\bz_i] = \bGamma \bz_i + \bzeta_i,\qquad
\Expectation[\bz_i \bzeta_i^\Trans]=\mathbf{0},
\end{align}
where $\bzeta_i$ is the error from the best linear approximation of $\Expectation[\dot{\bm}(\bw_i, \mu_i, \btheta_0)|\bz_i]$ based on $\bz_i$. This approximation error will not be small in general, because our paper allows for generic high-dimensional first-step covariates. However, in some special cases it can be small as we discuss further below.

The following assumption collects the key restrictions we impose on the first-step procedure.

\begin{Assumption}[First-Step Covariates]\label{Assumption: first-step}  \ \\
	\indent (i) $\max_{1\leq i\leq n}|\hat{\mu}_i-\mu_i|=\op(1)$.
	\\
	\indent (ii) $\Expectation[|\eta_i|^2]=o(n^{-1/2})$ and $\Expectation[|\eta_i|^2]\Expectation[|\bzeta_i|^2]=o(n^{-1})$.
\end{Assumption}

This assumption imposes high-level conditions on the covariates $\bz_i$ entering the first-step estimate (\ref{eq: first step, estimation}), covering both series-based nonparametric estimation and, more generally, many covariates settings. Assumption \ref{Assumption: first-step}(i) requires uniform consistency of $\hat{\mu}_i$ for $\mu_i$ only, without a convergence rate. In Section SA-2 of the Supplemental Appendix we discuss primitive conditions in different scenarios, covering (i) nonparametric series-based methods \citep{Belloni-Chernozhukov-Chetverikov-Kato_2015_JoE,Cattaneo-Farrell-Feng_2018_wp}, (ii) generic covariates with alternative conditions on the tails of their distribution, and (iii) generic covariates formed using many dummy/discrete regressors. The assumption also holds easily when covariates are discrete and a fully saturated model is used. This list is not meant to be exhaustive, and primitive conditions for other cases can be found in the vast literatures on nonparametric sieve estimation and high-dimensional models. Underlying this assumption is the implicit requirement that the best linear approximation of $\mu_i$ based on $\bz_i$ in (\ref{eq: first step, population}) should vanish asymptotically.

Assumption \ref{Assumption: first-step}(ii) concerns the approximation power of the covariates $\bz_i$ explicitly, measured in terms of the mean squared error of best linear approximations. It requires, at least, that the best linear approximation error in (\ref{eq: first step, population}) is sufficiently small relative to the sample size in mean square. The condition $\Expectation[|\eta_i|^2]=o(n^{-1/2})$ cannot be dropped without affecting the interpretation of the final estimand $\btheta_0$ because the first-step best linear approximation error will affect (in general) the probability limit of the resulting two-step estimator. In other words, either the researcher assumes that the best linear approximation is approximately exact in large samples, or needs to change the interpretation of the probability limit of the two-step estimator because of the misspecification introduced in the first step. The latter approach is common in empirical work, where researchers often employ a ``flexible'' parametric model, such as linear regression, Probit or Logit, all of which are misspecified in general.

Furthermore, the exact quality of approximation for the first-step estimate required in Assumption \ref{Assumption: first-step}(ii) depends on the quality of approximation in (\ref{eq: other first step, population}). At one extreme, the covariates $\bz_i$ may not offer any approximation of $\Expectation[\dot{\bm}(\bw_i, \mu, \btheta_0)|\bz_i]$ in mean square, in which case $\Expectation[|\bzeta_i|^2]=O(1)$, and hence the relevant restriction becomes $\Expectation[|\eta_i|^2]=o(n^{-1})$. This corresponds to the case of many generic covariates $\bz_i$ and non-linear $\Expectation[\dot{\bm}(\bw_i, \mu, \btheta_0)|\bz_i]$, that is, cases where $\bz_i$ are not basis of approximation and/or $\Expectation[\dot{\bm}(\bw_i, \mu, \btheta_0)|\bz_i]$ can not be well approximated by a linear combination of $\bz_i$.

At the other extreme, if $\Expectation[\dot{\bm}(\bw_i, {\mu}_i, \btheta_0) |\bz_i]$ can be well approximated by the best linear mean square prediction based on $\bz_i$ so that, at least, $\Expectation[|\bzeta_i|^2]=O(n^{-1/2})$, then the relevant restriction on the first-step estimate becomes $\Expectation[|\eta_i|^2]=o(n^{-1/2})$. This case encompasses the standard two-step semiparametric setup, where the covariates $\bz_i$ include basis expansions able to approximate $\mu_i=\Expectation[r_i|\bz_i]$ and $\Expectation[\dot{\bm}(\bw_i, \mu, \btheta_0)|\bz_i]$ accurately enough in mean square (usually justified by smoothness of these conditional expectations). From this perspective, the sufficient conditions $\Expectation[|\eta_i|^2]=o(n^{-1/2})$ and $\Expectation[|\bzeta_i|^2]=O(n^{-1/2})$ reassemble the usual requirement of better than $n^{1/4}$-consistency of first-step nonparametric estimators in two-step semiparametrics \citep[see][and references therein]{Cattaneo-Jansson_2018_ECMA}, but this is imposed only on best linear approximation errors (i.e., misspecification/smoothing bias), which are exacerbated for small $k$ and not for large $k$, the latter being the main focus of the present paper.

\begin{remark}[Extensions] In the Supplemental Appendix, we extend our main results in three directions. First, in Section SA-4.1 we allow for a multidimensional first-step $\bmu_i$ entering the second-step estimating equation $\bm(\bw,\cdot,\btheta)$. Second, in Section SA-4.2 we allow for a partially linear first-step structure as opposed to \eqref{eq: first step, population}. Both extensions are conceptually straightforward (they require additional notation and tedious algebra), but are nonetheless key to handle the production function example discussed in Section \ref{section: Olley-Pakes}. Finally, in Section SA-4.3 we discuss a special case of two-step estimators where high-dimensional covariates enter both the first-step (through $\mu_i$) and the second-step (in an additively separable way). This extension is useful in the context of marginal treatment effect estimation and inference, as we illustrate in Section \ref{section: MTE}. Allowing for the high-dimensional covariates to enter the second-step estimating equation in an unrestricted way makes the problem quite difficult, and therefore we relegate the general case for future work.
\end{remark}

\subsection{Distribution Theory}

It is not difficult to establish $\hat\btheta\to_\mathbb{P}\btheta_0$, even when $k/\sqrt{n}=O(1)$. See the Supplemental Appendix for exact regularity conditions. On the other hand, the $\sqrt{n}$-scaled mean squared error and distributional properties of the estimator $\hat\btheta$ will change depending on whether $k$ is ``small'' or ``larger'' relative to the sample size. To describe heuristically the result, consistency of $\hat\btheta$ and a second-order Taylor series expansion give:
\begin{align}
\label{eq: term 1}
\sqrt{n}(\hat\btheta-\btheta_0)  \approx&\ \frac{1}{\sqrt{n}}\bSigma_0\sum_{i=1}^{n} \bm(\bw_i, {\mu}_i, \btheta_0)\\
\label{eq: term 2}
&+ \frac{1}{\sqrt{n}}\bSigma_0\sum_{i=1}^{n} \dot{\bm}(\bw_i, {\mu}_i, \btheta_0)\Big(\hat{\mu}_i - \mu_i\Big)\\
\label{eq: term 3}
&+ \frac{1}{\sqrt{n}}\bSigma_0\sum_{i=1}^{n} \frac{1}{2}\ddot{\bm}(\bw_i, \mu_i, \btheta_0)\Big(\hat{\mu}_i - \mu_i\Big)^2,
\end{align}
where $\bSigma_0 =-(\bM_0^\Trans \boldsymbol{\Omega}_0\bM_0)^{-1}\bM_0^\Trans \boldsymbol{\Omega}_0$.

Term (\ref{eq: term 1}) will be part of the influence function. Using conventional large sample approximations (i.e., $k$ fixed or at most $k/\sqrt{n}\to0$), term (\ref{eq: term 2}) contributes to the variability of $\hat\btheta$ as a result of estimating the first step, and term (\ref{eq: term 3}) will be negligible. Here, however, we show that under the many covariates assumption $k/\sqrt{n}\not\to0$, both (\ref{eq: term 2}) and (\ref{eq: term 3}) will deliver nonvanishing bias terms. The main intuition is as follows: as the number of covariates increases relative to the sample size, the error in $\hat{\mu}_i-\mu_i$ also increases and features in terms (\ref{eq: term 2}) and (\ref{eq: term 3}). This, in turn, affects the finite sample performance of the usual asymptotic approximations, delivering unreliable results in applications. To be specific, the term (\ref{eq: term 2}) contributes a leave-in bias arising from using the same observation to estimate $\mu_i$ and later the parameter $\btheta_0$, while the term (\ref{eq: term 3}) contributes with a bias arising from averaging (non-linear) squared errors in the estimation of $\mu_i$.

The following theorem formalizes our main finding. The proof relies on several preliminary results given in the Supplemental Appendix. Let $\bZ=[\bz_1,\bz_2,\cdots,\bz_n]^\Trans$ be the first step included covariates and $\bPi=\bZ(\bZ^\Trans\bZ)^-\bZ^\Trans$ be the projection matrix with elements $\{\pi_{ij}:1\leq i,j\leq n\}$.

\begin{thm}[Asymptotic Normality]\label{Theorem: Asymptotic Representation}\ \\
Suppose Assumptions \ref{Assumption: first-step} and \ref{Assumption 2: Smoothness and Moments} hold, $\hat{\btheta} \to_\mathbb{P}\btheta_0$ the unique solution of $\Expectation[\bm(\bw_i,\mu_i,\btheta)]=\mathbf{0}$ and an interior point of $\Theta$, $\boldsymbol{\Omega}_n\to_\Prob \boldsymbol{\Omega}_0$ positive definite. If $k=O(\sqrt{n})$, then (\ref{eq:Result-1}) holds with
\[
\mathscr{B} = \bSigma_0\frac{1}{n}\sum_{i=1}^{n} \Expectation[\mathbf{B}_i|\bZ],\qquad
\mathscr{V} = \frac{1}{n}\bSigma_0\left(\mathbb{V}[\mathbb{E}[\bm(\bw_i,\mu_i,\btheta_0)|\bZ]] + \frac{1}{n}\sum_{i=1}^{n} \mathbb{V}[\bPsi_i|\bZ]\right)\bSigma_0,
\]
where
\[
\mathbf{B}_{i} = \dot{\bm}(\bw_i, {\mu}_i, \btheta_0)(r_i-\mu_i) \pi_{ii} +
                 \frac{1}{2}\ddot{\bm}(\bw_i,\mu_i,\btheta_0)\sum_{j=1}^{n}(r_j-\mu_j)^2\pi_{ij}^2,
\]
\[
\bPsi_i = \bm(\bw_i,\mu_i,\btheta_0) +
          \left(\sum_{j=1}^{n}\Expectation[\dot{\bm}(\bw_j, {\mu}_j, \btheta_0) |\bZ]\pi_{ij}\right)(r_i-\mu_i).
\]
\end{thm}

Using well known properties of projection matrices, it follows that $\mathscr{B}=O_\mathbb{P}(k/n)$ and non-zero in general, and thus the distributional approximation in Theorem \ref{Theorem: Asymptotic Representation} will exhibit a first-order asymptotic bias $\mathscr{V}^{-1/2}\mathscr{B}$ whenever $k$ is ``large'' relative to the sample size (e.g., $k\propto\sqrt{n}$). In turn, this result implies that conventional inference procedures ignoring this first-order distributional bias will be invalid, leading to over-rejection of the null hypothesis of interest and under-coverage of the associated confidence intervals. Section \ref{section: numerical} presents simulation evidence capturing this phenomena.

To understand the implications of the above theorem, we discuss the two terms in $\mathbf{B}_{i}$. The first term corresponds to the contribution from (\ref{eq: term 2}), because a first order approximation gives ${\bm}(\bw_i, \hat{\mu}_i, \btheta_0)\approx \dot{\bm}(\bw_i, {\mu}_i, \btheta_0)(\hat{\mu}_i-\mu_i)\approx \dot{\bm}(\bw_i, {\mu}_i, \btheta_0)(\sum_j \pi_{ij}(r_j-\mu_j))$. Because $\mathbb{E}[r_j-\mu_j|\bz_j]=0$, this bias is proportional to the sample average of $\Cov[\dot{\bm}(\bw_i, {\mu}_i, \btheta_0), r_i-\mu_i|\bz_i]\pi_{ii}$. Hence the bias, due to the linear contribution of $\hat{\mu}_i$, will be zero if there is no residual variation in the sensitivity measure $\dot{\bm}$ (i.e., $\Var[\dot{\bm}(\bw_i, {\mu}_i, \btheta_0) |\bz_i]=\mathbf{0}$) or, more generally, the residual variation in the sensitivity measure $\dot{\bm}$ is uncorrelated to the first step error term (i.e., $\Cov[\dot{\bm}(\bw_i, {\mu}_i, \btheta_0),r_i-\mu_i |\bz_i]=\mathbf{0}$).

The second term in $\mathbf{B}_{i}$ captures the quadratic dependence of the estimating equation on the unobserved $\mu_i$, coming from (\ref{eq: term 3}). Because of the quadratic nature, this bias represents the accumulated estimation error when $\hat{\mu}_i$ is overfitted. When $i\neq j$, which is the main part of the bias, $\Expectation[ \ddot{\bm}(\bw_i,\mu_i,\btheta_0)(r_j-\mu_j)^2| \bz_i, \bz_j] = \Expectation[ \ddot{\bm}(\bw_i,\mu_i,\btheta_0) | \bz_i]\Expectation[ (r_j-\mu_j)^2 | \bz_j]$, and hence this portion of the bias will be non-zero unless an estimating equation linear in $\mu_i$ is considered or, slightly more generally, $\Expectation[\ddot{\bm}(\bw_i,\mu_i,\btheta_0) | \bz_i]=\mathbf{0}$. Intuitively, overfitting the first step does not give a quadratic contribution if the estimating equation is not sensitive to the first step on average to the second order.

The first bias can be manually removed by employing a leave-one-out estimator of $\mu_i$. However, the second bias cannot be removed this way. Furthermore, the leave-one-out estimator $\hat{\mu}_i^{(i)}$ usually has higher variability compared with $\hat{\mu}_i$, hence the second bias will be amplified, which is confirmed by our simulations. 

\citet{Chernozhukov-Escanciano-Ichimura-Newey-Robins_2018_LocRobust} introduced the class of locally robust estimators, which are a generalization of doubly robust estimators \citep{Bang-Robins_2005_Biometrics} and the efficient influence function estimators \citep[p. 142]{Cattaneo_2010_JoE}. These estimators can offer demonstrable improvements in terms of smoothing/approximation bias rate restrictions and, consequently, they offer robustness to ``small'' $k$ (underfitting). See also \citet{Chernozhukov-Chetverikov-Demirer-Duflo-Hansen-Newey-Robins_2018_EJ} and \citet{Newey-Robins_2018_wp} for related approaches. This type of estimators are carefully constructed so that \eqref{eq: term 2} is removed, but they do not account for \eqref{eq: term 3}.  Because the ``large'' $k$ bias is in part characterized by \eqref{eq: term 3}, locally robust estimators cannot (in general) reduce the bias we uncover in this paper. Therefore, our methods complement locally robust estimation by offering robustness to overfitting, that is, situations where the first step estimate includes possibly many covariates.
\citet{Cattaneo-Jansson_2018_ECMA} illustrate this fact in the context of kernel-based estimation.

Consider next the variance and distributional approximation. Theorem \ref{Theorem: Asymptotic Representation} shows that the distributional properties of $\hat{\btheta}$ are based on a double sum in general, and hence it does not have an ``influence function'' or asymptotically linear representation. Nevertheless, after proper Studentization, asymptotic normality holds as in (\ref{eq:Result-1}). The following remark summarizes the special case when the estimator, after bias correction, does have an asymptotic linear representation.

\begin{remark}[Asymptotic Linear Representation] Suppose the conditions of Theorem \ref{Theorem: Asymptotic Representation} hold. If, in addition, $\Expectation[|\bzeta_i|^2]=o(1)$, then
\[
\sqrt{n}(\hat{\btheta} - \btheta_0 - \mathscr{B})
= \bSigma_0\frac{1}{\sqrt{n}}\sum_{i=1}^n \Big\{\bm(\bw_i,\mu_i,\btheta_0) 
	                              + \Expectation[\dot{\bm}(\bw_i, {\mu}_i, \btheta_0) |\bz_i](r_i-\mu_i)\Big\} + \op(1),
\]
hence $\hat{\btheta}$ is asymptotically linear after bias correction even when $k/\sqrt{n}\not\to0$. However, $\hat{\btheta}$ is asymptotically linear if and only if $k/\sqrt{n}\to0$ in general. See \citet{Newey_1994_ECMA} and \citet{Hahn-Ridder_2013_ECMA} for more discussion on asymptotic linearity and variance calculations.
\end{remark}

In practice one needs to estimate both the bias and the variance to conduct valid statistical inference. Plug-in estimators could be constructed to this end, though additional unknown functions would need to be estimated (e.g., conditional expectations of derivatives of the estimating equation). Under regularity conditions, these estimators would be consistent for the bias and variance terms. As a practically relevant alternative, we show in the upcoming sections that the jackknife can be used to estimate both the bias and variance, and that a carefully crafted resampling method can be used to conduct inference. The key advantage of these results is that they are fully automatic, and therefore can be used for any model considered in practice without having to re-derive and plug-in for the exact expressions each time. 

\begin{remark}[Delta Method]\label{rmk: delta method}
Our results apply directly to many other estimands via the so-called delta method. Let $\boldsymbol{\varphi}(\cdot)$ be a possibly vector-valued continuously differentiable function of the parameter $\btheta_0$ with gradient $\dot{\boldsymbol{\varphi}}(\cdot)$. Then, under the conditions of Theorem \ref{Theorem: Asymptotic Representation},
\[
\Big(\dot{\boldsymbol{\varphi}}(\btheta_0)\mathscr{V}\dot{\boldsymbol{\varphi}}(\btheta_0)^{\Trans}\Big)^{-1/2}
\Big(\boldsymbol{\varphi}(\hat{\btheta})-\boldsymbol{\varphi}(\btheta_0) - \dot{\boldsymbol{\varphi}}(\btheta_0)\mathscr{B}\Big)
\ \rightsquigarrow\ \mathrm{Normal}(\mathbf{0},\ \mathbf{I}),
\]
provided that $\dot{\boldsymbol{\varphi}}(\btheta_0)$ is full rank. Hence, the usual delta method can be used for estimation and inference in our setting, despite the presence of potentially many covariates entering the first-step estimate.
\end{remark}

Plug-in consistent estimation of the appropriate GMM efficient weighting matrix is also possible given our regularity conditions, but we do not give details here to conserve space.

\section{Jackknife Bias Correction and Variance Estimation}\label{section: jackknife}

We show that the jackknife is able to estimate consistently the many covariate bias and the asymptotic variance of $\hat{\btheta}$, even when $k=O(\sqrt{n})$, and without assuming a valid asymptotic linear representation for $\hat{\btheta}$.

The jackknife estimates are constructed by simply deleting one observation at the time and then re-estimating both the first and second steps. To be more specific, let $\hat{\mu}^{(\ell)}_{i}$ denote the first-step estimate after the $\ell^{\text{th}}$ observation is removed from the dataset. Then, the leave-$\ell$-out two-step estimator is 
\[\hat{\btheta}^{(\ell)}
  = \arg\min_{\btheta} \left|\boldsymbol{\Omega}_n^{1/2}\sum_{i=1, i\neq \ell}^{n} \bm(\bw_i, \hat{\mu}_i^{(\ell)},{\btheta})\right|,\qquad \ell=1,2,\dots,n.
\]
Finally, the bias and variance estimates are constructed as in (\ref{eq:Bias-Variance}). This approach is fully data-driven and automatic. In addition, another appealing feature of the jackknife in our case is that it is possible to exploit the specific structure of the problem to reduce computational burden. Specifically, because we consider a linear regression fit for the first step, the leave-$\ell$-out estimate $\hat{\mu}^{(\ell)}_{i}$ can easily be obtained by
\begin{align*}
\hat{\mu}^{(\ell)}_{i} &= \hat{\mu}_{i}+ \frac{\hat{\mu}_{\ell}-  r_\ell}{1-\pi_{\ell\ell}} \cdot\pi_{i\ell},\qquad 1\leq i\leq n,
\end{align*}
where recall that $\pi_{i\ell}$ is the $(i,\ell)^{\text{th}}$ element of the projection matrix for the first step $\bPi=\bZ(\bZ^\Trans\bZ)^-\bZ^\Trans$. Since recomputing the first-step estimate can be time-consuming when $k$ is large, the above greatly simplifies the algorithm and reduces computing time. 

To show the validity of the jackknife, we impose two additional mild assumptions on the possibly large dimensional covariates $\bz_i$, captured through the projection matrix of the first-step estimate.

\begin{thm}[Jackknife-Based Valid Inference]\label{Theorem: Jackknife}\ \\
Suppose the conditions of Theorem \ref{Theorem: Asymptotic Representation} hold. If, in addition, (i) $\sum_{1\leq i\leq n} \pi_{ii}^2=\op(k)$, (ii) $\max_{1\leq i\leq n} 1/(1-\pi_{ii})=\Op(1)$ and (iii) $\Expectation[\epsilon_i^6|\bz_i]$ is uniformly bounded, then (\ref{eq:Result-2}) holds.
\end{thm} 

The two conditions together correspond to ``design balance'', which states that asymptotically the projection matrix is not ``concentrated'' on a few observations. They are slightly weaker than $\max_{1\leq i\leq n}\pi_{ii}=\op(1)$, which is commonly assumed in the literature on high-dimensional statistics. In Section SA-2.4 of the Supplemental Appendix we give a concrete example using sparse dummy covariates where $\max_{1\leq i\leq n}\pi_{ii}\not=\op(1)$, but the conditions (i) and (ii) are satisfied. For more discussion on design balance in linear least squares models see, e.g., \citet{Chatterjee-Hadi_1988_Book}.

\begin{remark}[Delta Method]\label{rmk: delta method-2}
Consider the setup of Remark \ref{rmk: delta method}, where the goal is to conduct estimation and inference for a (smooth) function of $\btheta_0$. In this case, the estimator is $\boldsymbol{\varphi}(\hat{\btheta})$. There are at least three ways to conduct bias correction: (i) plug-in method leading to $\boldsymbol{\varphi}(\hat{\btheta} - \hat{\mathscr{B}})$, (ii) linearization-based method leading to $\boldsymbol{\varphi}(\hat{\btheta})-\dot{\boldsymbol{\varphi}}(\hat{\btheta})\hat{\mathscr{B}}$, and (iii) direct jackknife of $\boldsymbol{\varphi}(\hat{\btheta})$. The three methods are asymptotically equivalent, and can be easily implemented in practice. The same argument applies to the variance estimator when $\boldsymbol{\varphi}(\btheta_0)$ is the target parameter.
\end{remark}

By showing the validity of the jackknife, one can construct confidence intervals and conduct hypothesis tests using the jackknife bias and variance estimators, and the normal approximation. In particular, bias correction will not affect the variance of the asymptotic distribution. On the other hand, any bias correction technique is likely to introduce additional variability, which can be nontrivial in finite samples. This is indeed confirmed by our simulation studies. In the next section, we introduce a carefully crafted fully automatic bootstrap method that can be applied to the bias-corrected Studentized statistic to obtain better finite sample distributional approximations.

\section{Bootstrap Inference after Bias Correction}\label{section: bootstrap}

In this section we develop a fast, automatic and specifically tailored bootstrap-based approach to conducting post-bias-correction inference in our setting. The method combines the wild bootstrap (first-step estimation) and the multiplier bootstrap (second-step estimation) to give an easy-to-implement valid distributional approximation to the finite sample distribution of the jackknife-based bias-corrected Studentized statistic in (\ref{eq:Result-2}). See \citet{Mammen_1993_AoS} for a related result in the context of a high-dimensional one-step linear regression model without any bias-correction, and \citet{Kline-Santos_2012_JoE} for some recent higher-order results in the context of parametric low-dimensional linear regression models.

Let $\omega_i^\star$, $i=1,2,\cdots,n$ be i.i.d. bootstrap weights with $\Expectation[\omega_i^\star]=1$, $\mathbb{V}[\omega_i^\star]=1$, $\Expectation[(\omega_i^\star-1)^3]=0$ and finite fourth moment. First, we describe the bootstrap construction of $\hat{\btheta}^\star$. We employ the wild bootstrap to obtain $\hat{\mu}_i^\star$, mimicking the first-step estimate (\ref{eq: first step, estimation}): we regress $r_i^{\star}$ on $\bz_i$, where $r_i^\star = \hat{\mu}_i + (\omega_i^\star-1)(r_i-\hat{\mu}_i)$. Then, we employ the multiplier bootstrap to obtain $\hat{\btheta}^\star$, mimicking the second-step estimate (\ref{eq: second step, estimation}):
\begin{align}\label{eq:bootstrap theta}
\hat{\btheta}^\star = \arg\min_{\btheta} \left|\boldsymbol{\Omega}_n^{1/2}\sum_{i=1}^{n} \omega_i^\star\bm(\bw_i, \hat{\mu}_i^\star,{\btheta})\right|.
\end{align}
Second, we describe the bootstrap construction of $\hat{\mathscr{B}}$ and $\hat{\mathscr{V}}$; that is, the implementation of the jackknife bias and variance estimators under the bootstrap. Because we employ a multiplier bootstrap, the jackknife estimates need to be adjusted to account for the effective number of observations under the bootstrap law. Thus, we have:
\[\hat{\mathscr{B}}^{\star} = (n-1) (\hat{\btheta}^{\star,(\cdot)}- \hat{\btheta}^{\star}),\qquad
  \hat{\mathscr{V}}^{\star} = \frac{n-1}{n} \sum_{\ell=1}^n \omega_\ell^\star
	                            (\hat{\btheta}^{\star,(\ell)} - \hat{\btheta}^{\star,(\cdot)})
	                            (\hat{\btheta}^{\star,(\ell)} - \hat{\btheta}^{\star,(\cdot)})^\Trans,
\]
where
\[
\hat{\btheta}^{\star,(\cdot)} = \frac{1}{n}\sum_{\ell=1}^n\omega_\ell^\star\hat{\btheta}^{\star,(\ell)},\]
\[
\hat{\btheta}^{\star,(\ell)} = \arg\min_{\btheta}
	   \left|\boldsymbol{\Omega}_n^{1/2}\Big\{\sum_{i=1}^{n} \big[\omega_i^\star-\Indicator_{(i=\ell)}\big]\bm(\bw_i, \hat{\mu}_i^{\star,(\ell)},{\btheta})
		                                             \Big\}\right|,\qquad
	\ell = 1,2,\dots,n.
\]
Here $\hat{\mu}_i^{\star,(\ell)}$ is obtained by regressing $r_i^\star$ on $\bz_i$, without using the $\ell^{\text{th}}$ observation. Equivalently, the jackknife deletes the $\ell^{\text{th}}$ observation in the first step wild bootstrap, and reduces the $\ell^{\text{th}}$ weight $\omega_\ell^\star$ by 1 in the second step multiplier bootstrap.

Our resampling approach employs the wild bootstrap to form $\hat{\mu}_i^\star$, which is very easy and fast to implement and does not require recomputing the possibly high-dimensional projection matrix $\bPi$, and then uses the same bootstrap weights to construct $\hat{\btheta}^\star$ via a multiplier resampling approach. It is possible to use the multiplier bootstrap for both estimation steps, which would give a more unified treatment, but such an approach is harder to implement and does not utilize efficiently (from a computational point of view) the specific structure of the first-step estimate. To be more specific, employing the multiplier bootstrap in the first-step estimation leads to $\hat{\mu}_i^\star = \bz_i^\Trans (\bZ^\Trans \mathbf{W}^\star \bZ)^- \bZ^\Trans \mathbf{W}^\star \bR$, where $\bR=[r_1,r_2,\dots,r_n]^\Trans$ and $\mathbf{W}^\star$ is a diagonal matrix with diagonal elements $\{\omega_i^\star\}_{1\leq i\leq n}$, which requires recomputing the projection matrix for each bootstrap replication. In contrast, our bootstrap approach leads to $\hat{\mu}_i^\star = \bz_i^\Trans (\bZ^\Trans\bZ)^- \bZ^\Trans \bR^\star$, where $\bR^\star=[r^\star_1,r^\star_2,\dots,r^\star_n]^\Trans$. As discussed before, this important practical simplification also occurs because we are employing a linear regression fit in the first step. Employing the standard nonparametric bootstrap may also be possible, but additional (stronger) regularity conditions would be required. Last but not least, we note that combining the jackknife with the multiplier bootstrap na\"ively (that is, deleting the $\ell^{\text{th}}$ observation with its weight $\omega_\ell^\star$ altogether in the second step) does not deliver a consistent variance estimate; see the Supplemental Appendix for details.

The following theorem summarizes our main result for inference. Only two additional mild, high-level conditions on the bootstrap analogue first-step and second-step estimators are imposed.

\begin{thm}[Bootstrap Validity]\label{Theorem: bootstrap+jackknife}\ \\
Suppose the conditions of Theorems \ref{Theorem: Asymptotic Representation} and \ref{Theorem: Jackknife} hold. If, in addition, $\max_{1\leq i\leq n}|\hat{\mu}_i^\star - \hat{\mu}_i|=\op(1)$ and $|\hat{\btheta}^\star - \hat{\btheta}| =\op(1)$, then (\ref{eq:Result-3}) holds.
\end{thm}

It is common to assume the bootstrap weights $\omega^\star_i$ to have mean 1 and variance 1. For the jackknife bias and variance estimator to be consistent under the bootstrap distribution, we also need that the third central moment of $\omega^\star_i$ is zero. Examples include $\omega^\star_i = 1+e_i^\star$ with $e_i^\star$ following the Rademacher distribution or the six-point distribution proposed in \cite{Webb_2014_wp}.

For inference, consider for example the one dimensional case: $\dim(\btheta_0)=1$. The bootstrap percentile-t bias-corrected (equal tail) confidence interval for $\theta_0$ is
\[
  \left[\;\hat{\theta} - \hat{\mathscr{B}} - \hat{q}_{1-\alpha/2} \cdot \sqrt{\hat{\mathscr{V}}}
	        \quad,\quad
	        \hat{\theta} - \hat{\mathscr{B}} - \hat{q}_{\alpha/2}   \cdot \sqrt{\hat{\mathscr{V}}}\;
	\right], 
\]
where $\hat{q}_{\alpha} = \inf\{t\in\mathbb{R}:\hat{F}(t)\geq \alpha\}$ is the empirical $\alpha^{\text{th}}$ quantile of $\{\mathscr{T}^\star_b : 1\leq b\leq B \}$, with $\hat{F}(t) = \frac{1}{B} \sum_{b=1}^B \Indicator[\mathscr{T}^\star_b\leq t]$ and $\mathscr{T}^\star_b$ denoting the bootstrap statistic in (\ref{eq:Result-3}) in $b^{\text{th}}$ simulation.

\section{Examples}\label{section: examples}

We now apply our results to two examples: production function estimation, which is an IO application with multidimensional partially linear first step estimation, and marginal treatment effect estimation, which relates to IV methods with heterogeneous treatment effects and many covariates/instruments. The Supplemental Appendix analyzes several other examples in applied microeconomics. For each example, besides the general form of the bias in Theorem \ref{Theorem: Asymptotic Representation}, we show that it is possible to further characterize the nature and source of the many covariates bias by utilizing corresponding identification assumptions in each of the examples.

\subsection{Production Function}\label{section: Olley-Pakes}

As a first substantive application of our results to empirically relevant problems in applied microeconomics, we consider production function estimation. For a review of this topic, including an in-depth discussion of its applicability to industrial organization and other fields in Economics, see \citet{Ackerberg-Benkard-Berry-Pakes_2007_Handbook}. For concreteness, here we focus on the setting introduced by \cite{Olley-Pakes-1996_ECMA}, and propose new estimation and inference methods for production functions allowing for possibly many covariates in the first step estimation. To apply our methods to this problem, two extensions mentioned previously (multidimensional and partially linear first step estimation) are needed, and therefore the results below are notationally more involved.

We use $i$ to index firms (i.e., observations) and $t$ for time. The production function takes the form:
\begin{align}\label{eq:output1}
Y_{i,t} &= \beta_L L_{i,t}  + \beta_K K_{i,t} + \beta_A A_{i,t} + W_{i,t} + U_{i,t}, 
\end{align} 
where $Y$, $L$, $K$ and $A$ represent (log) output, labor input, capital input and aging effect, respectively. $W$ is the firm-specific productivity factor, and is a (generalized) fixed effect. The error term $U$ is either measurement error or shock that is unpredictable with time-$t$ information, hence has zero conditional mean (given the right-hand-side variables). Since the productivity factor $W$ is unobserved, (\ref{eq:output1}) cannot be used directly to estimate the production function. 

Now we discuss briefly the decision process of a firm in each period. First, the firm compares continuation value to salvage (liquidation) value, and decides whether or not to exit the market. Upon deciding to stay in business, the firm chooses simultaneously the labor input $L_{i,t}$ and investment $I_{i,t}$, given its private knowledge about productivity $W_{i,t}$. Finally, capital stock follows the classical law of motion.

The first crucial assumption for identification is that there is a one-to-one relationship between the firm-level decision variable $I_{i,t}$ and the unobserved state variable $W_{i,t}$, which allows inverting the investment decision and writing $W_{i,t} = h_t(I_{i,t}, K_{i,t}, A_{i,t})$, with $h_t$ unknown and possibly time-dependent. Then, (\ref{eq:output1}) can be written as
\begin{align}\label{eq:output2}
Y_{i,t} &= \beta_L L_{i,t} + \phi_{i,t} + U_{i,t},\qquad \phi_{i,t}=\beta_K K_{i,t} + \beta_A A_{i,t}+h_t(I_{i,t}, K_{i,t}, A_{i,t}).
\end{align}
The above equation can be used to estimate the labor share $\beta_L$ flexibly using a partially linear regression approach. The capital share $\beta_K$ and the effect of aging $\beta_A$, however, are not identified. 

To identify $\beta_K$ and $\beta_A$, \cite{Olley-Pakes-1996_ECMA} use information embedded in the firm's exit decision, which is shown as $\chi_{i,t}=\Indicator[W_{i,t}\geq\underline{W}_t(A_{i,t},K_{i,t})]$, where $\chi_{i,t}=1$ represents the firm staying in business, and $\underline{W}_t$ is the threshold function. Then, we decompose $W_{i,t+1}$ into 
\[W_{i,t+1} = \Expectation[W_{i,t+1}|W_{i,t},\chi_{i,t+1}=1] + V_{i,t+1}.\]
Conditioning on survival at time $t+1$ (i.e. $\chi_{i,t+1}=1$) is the same as conditioning on $\underline{W}_{i,t}$, and hence the conditional expectation in the above display is an unknown function of $W_{i,t}$ and $\underline{W}_{i,t}$. The second crucial assumption for identification is that the survival probability, defined as
\begin{align}\label{eq:prob of stay}
P_{i,t} &= P_t(I_{i,t},K_{i,t},A_{i,t})=\Expectation[\chi_{i,t+1}|I_{i,t},K_{i,t},A_{i,t}],
\end{align}
is a valid proxy for $\underline{W}_{i,t}$. Therefore, with the time index progressed by one period, we rewrite (\ref{eq:output1}) as
\begin{alignat}{2}
\nonumber Y_{i,t+1} - \beta_L L_{i,t+1} &= \beta_K K_{i,t+1} + \beta_A A_{i,t+1} + g(P_{i,t},\ W_{i,t}) &&+ V_{i,t+1} + U_{i,t+1}\\
\label{eq:output3} &= \beta_K K_{i,t+1} + \beta_A A_{i,t+1} + g(P_{i,t},\ \phi_{i,t}-\beta_K K_{i,t} - \beta_A A_{i,t}) &&+ V_{i,t+1} + U_{i,t+1}.
\end{alignat}
Here we make an important remark on the two error terms and why labor input has been moved to the left-hand-side, which also sheds light on the estimation strategy. $U_{i,t+1}$ is either a measurement error or the conditional expectation error of $Y_{i,t+1}$ on contemporaneous variables, hence is orthogonal to time-$t+1$ information. On the other hand, $V_{i,t+1}$ is the conditional expectation error of $\chi_{i,t+1}$ on time-$t$ variables, hence is only orthogonal to time-$t$ information. It is uncorrelated with $K_{i,t+1}$ and $A_{i,t+1}$ since they are predetermined, but in general \textit{correlated}
with $L_{i,t+1}$. This is the endogeneity problem underlying (\ref{eq:output3}), and shows why it cannot be used for estimation without $\beta_L$ being estimated in a first step.

Now we describe the estimation strategy. For simplicity we make two assumptions: (i) there are only two periods $t=1,2$, and (ii) the function $g$ is known up to a finite dimensional parameter $\blambda_0$. First, we rely on (\ref{eq:output2}) to estimate $\beta_L$ and $\phi_{i,1}$ with a partially linear regression, which gives $\hat{\beta}_L$ and $\hat{\phi}_{i,1}$. Second, we use (\ref{eq:prob of stay}) to obtain the estimated probability of staying, $\hat{P}_{i,1}$. These are the two first-step estimates in this application. Finally, given the preliminary estimates, $\beta_K$, $\beta_L$ and the nuisance parameter $\blambda_0$ are jointly estimated in the second step. The entire two-step estimation approach is summarized as follows:
\begin{align*}
\left[\begin{array}{c}\hat\beta_K\\\hat\beta_A\\\hat\blambda\end{array}\right]
&=\argmin_{\beta_K,\beta_A,\blambda} \frac{1}{n}\sum_{i=1}^n  \Big[Y_{i,2} - \hat{\beta}_L L_{i,2} - \beta_K K_{i,2} - \beta_A A_{i,2} - g(\hat{P}_{i,1}, \hat{\phi}_{i,1} - \beta_K K_{i,1} - \beta_A A_{i,1}, \blambda)\Big]^2,\\
\hat{\phi}_{i,1}
&= \bZ_{i,1}^\Trans\hat{\bgamma}_1,\qquad [\hat{\beta}_L,\hat{\bgamma}_1^\Trans]^\Trans = \argmin_{\beta,\bgamma}\sum_{i=1}^{n}\Big( Y_{i,1} - \beta L_{i,1} - \bZ_{i,1}^\Trans\bgamma \Big)^2,\\
\hat{P}_{i,1} 
&= \bZ_{i,1}^\Trans\hat{\bgamma}_2,\qquad \hat{\bgamma}_2 = \argmin_{\bgamma}\sum_{i=1}^n\Big(\chi_{i,2} - \bZ_{i,1}^\Trans\bgamma\Big)^2,
\end{align*}   
with $\bZ_{i,1}$ being series expansion based on the variables $(I_{i,1},K_{i,1},A_{i,1})$, in addition to perhaps other variables.

The estimation problem does not fit into our basic framework for three reasons. First, we have two estimating equations in the first step, one for $(\beta_L,\phi_{i,1})$ and the other for $P_{i,1}$. Second, the model features a parameter ($\beta_L$) estimated in the first step and then plugged into the second step estimating equation. Third, $\hat{\phi}_{i,1}$ is no longer a conditional expectation projection, but is instead obtained from a partially linear regression. As mentioned above, Section SA-4 in the Supplemental Appendix discusses extensions of our framework that enable us to handle this application in full generality. 

Applying Theorem \ref{Theorem: Asymptotic Representation}, properly extended using the results in Section SA-4 of the Supplemental Appendix, we have the following results on bias and variance for the estimator $[\hat{\beta}_K,\hat{\beta}_A,\hat{\blambda}^\Trans]^\Trans$.

\begin{coro}[Asymptotic Normality: Production Function]\ \\
Assume the assumptions of Theorem \ref{Theorem: Asymptotic Representation} and the example-specific additional regularity conditions summarized in the Supplemental Appendix hold. Then,
\begin{align*}
\mathbf{B}_{i} &= \Big(\mathbf{b}_{1,i}+\mathbf{b}_{2,i}\Big)\pi_{ii} + \sum_{j=1}^n\Big(\mathbf{b}_{3,ij} + \mathbf{b}_{4,ij}+ \mathbf{b}_{5,ij}\Big)\pi_{ij}^2\\
\boldsymbol{\Psi}_i &= \boldsymbol{\Psi}_{1,i} + \boldsymbol{\Psi}_{2,i} + \boldsymbol{\Psi}_{3,i},
\end{align*}
where
\begin{align*}
\bb_{1,i} &= \begin{bmatrix}
K_{i,1} g_{22,i} \\
A_{i,1} g_{22, i}\\
-\bg_{23,i}  
\end{bmatrix}U_{i,1}V_{i,2},\qquad
\bb_{2,i} = \begin{bmatrix}
K_{i,1} g_{12,i} \\
A_{i,1} g_{12, i}\\
-\bg_{13,i}  
\end{bmatrix}(\chi_{i,2}-P_{i,1})V_{i,2}\\
\bb_{3,ij} &= \left( -\begin{bmatrix}
K_{i,1} g_{22,i} \\
A_{i,1} g_{22, i}\\
-\bg_{23,i}  
\end{bmatrix}g_{2,i}  -\frac{1}{2} \begin{bmatrix}
K_{i,1} g_{2,i} - K_{i,2}  \\
A_{i,1} g_{2,i} - A_{i,2}\\
-\bg_{3,i} 
\end{bmatrix}g_{22,i} \right)U_{i,1}^2,\\
\bb_{4,ij} &= \left(-\begin{bmatrix}
                      K_{i,1} g_{12,i} \\
                      A_{i,1} g_{12, i}\\
                      -\bg_{13,i}
                      \end{bmatrix}g_{1,i}
                    -\frac{1}{2} \begin{bmatrix}
                                 K_{i,1} g_{2,i} - K_{i,2}  \\
                                 A_{i,1} g_{2,i} - A_{i,2}\\
                                 -\bg_{3,i} 
                                 \end{bmatrix}g_{11,i} \right)(\chi_{j,2}-P_{j,1})^2,\\
\bb_{5,ij} &=\left(
\begin{bmatrix}
K_{i,1} g_{22,i} \\
A_{i,1} g_{22, i}\\
-\bg_{23,i}  
\end{bmatrix}g_{1,i}
+ \begin{bmatrix}
K_{i,1} g_{12,i}   \\
A_{i,1} g_{12,i} \\
-\bg_{13,i} 
\end{bmatrix}g_{2,i}
+ \begin{bmatrix}
K_{i,1} g_{2,i} - K_{i,2}  \\
A_{i,1} g_{2,i} - A_{i,2}\\
-\bg_{3,i} 
\end{bmatrix}g_{12,i}
\right) (\chi_{j,2}-P_{j,1})U_{j,1}\\
\boldsymbol{\Psi}_{1,i} &= \begin{bmatrix}
K_{i,1} g_{2,i}-K_{i,2}\\
A_{i,1} g_{2,i}-A_{i,2}\\
-\bg_{3,i}
\end{bmatrix}\Big(V_{i,2} + U_{i,2}\Big)\\
\boldsymbol{\Psi}_{2,i} &= -\begin{bmatrix}
K_{i,1} g_{2,i} - K_{i,2}  \\
A_{i,1} g_{2,i} - A_{i,2}\\
-\bg_{3,i} 
\end{bmatrix}g_{2,i} U_{i,1} - \begin{bmatrix}
K_{i,1} g_{2,i} - K_{i,2}  \\
A_{i,1} g_{2,i} - A_{i,2}\\
-\bg_{3,i} 
\end{bmatrix}g_{1,i} \Big(\chi_{i,2} - P_{i,1}\Big)\\
\boldsymbol{\Psi}_{3,i} &= \frac{1}{\Expectation[\Var[L_{i,1}|\bZ_{i,1}]]}\boldsymbol{\Xi}_0 \Big(L_{i,1}-\Expectation[L_{i,1}\Big|\bZ_{i,1}]\Big) U_{i,1}.
\end{align*}
We use the abbreviation $g_i=g(P_{i,1}, W_{i,1},\blambda_0)$, and further subscripts $1$, $2$ and $3$ of $g_i$ are used to denote its partial derivatives with respect to the first, second and third argument, respectively. Exact formulas of $\bSigma_0$ and $\boldsymbol{\Xi}_0$ are available in the Supplemental Appendix (Section SA-4.2 and SA-5.7). 
\end{coro} 

Some bias terms can be made to zero with additional assumptions. First consider the scenario that $U_{i,t}$ is purely measurement error. Then it should be independent of other error terms, which implies $\bb_{1,i}$ and $\bb_{5,ij}$ are zero after taking conditional expectations. Sometimes it is assumed that all firms survive from time-$1$ to time-$2$ (i.e., there is no sample attrition), or the analyst focuses on a subsample (in which case the parameters have to be reinterpreted). Then $\chi_{i,2}=P_{i,1}=1$, hence $\bb_{2,i}$ and $\bb_{4,ij}$ are zero. The variance term $\boldsymbol{\Psi}_{i,2}$ also simplifies.

\subsection{Marginal Treatment Effect}\label{section: MTE}

Originally proposed by \citet{Bjorklund-Moffitt_1987_ReStat}, and later developed and popularized by \citet{Heckman-Vytlacil_2005_ECMA} and \citet{Heckman-Urzua-Vytlacil_2006_ReStat}, the marginal treatment effect (MTE) is an important parameter of interest in program evaluation and causal inference. Not only it can be viewed as a limiting version of the local average treatment effect (LATE) of \citet{Imbens-Angrist_1994_ECMA} for continuous instrumental variables \citep[c.f.][]{Angrist-Graddy-Imbens_2000_ReStud}, but also it can be used to unify and interpret many other treatment effects parameters such as the average treatment effect or the treatment effect on the treated. Another appealing feature of the MTE is that it provides a description of treatment effect heterogeneity.

To describe the MTE, we adopt a potential outcomes framework under random sampling. Suppose $(Y_i,T_i,\bX_i,\bZ_i)$, $i=1,2,\dots,n$, is i.i.d., where $Y_i$ is the outcome of interest, $T_i$ is a treatment status indicator, $\bX_i\in\mathbb{R}^{d_x}$ is a $d_x$-variate vector of observable characteristics, and $\bZ_i\in\mathbb{R}^{k}$ is $k$-variate vector of ``instruments'' (which may include $\bX_i$ and transformations thereof). The observed data is generated according to the following switching regression model, also known as potential outcomes or the Roy model,
\begin{alignat}{3}
\label{eq: switching regression 1} 
Y_i &= T_i Y_{i}(1) + (1-T_i) Y_{i}(0), &&\qquad Y_i(1) = g_1(\bX_i) + U_{1i},
&&\qquad  Y_i(0) = g_0(\bX_i) + U_{0i},\\
\label{eq: switching regression 2}
T_i &= \Indicator[P_i \geq V_i],    &&\qquad P_i=P(\bZ_i) = \Expectation[T_i|\bZ_i],
&&\qquad V_i|\bX_i\sim\mathrm{Uniform}[0,1],
\end{alignat}
where $Y_i(1)$ and $Y_i(0)$ are the potential outcomes when an individual receives the treatment or not, $(U_{1i},U_{0i},V_i)$ are unobserved error terms, and $P_i$ is the propensity score or probability of selection. The selection equation (\ref{eq: switching regression 2}) is taken essentially without loss of generality to be of the single threshold-crossing form \citep[see][for more discussion]{Vytlacil_2002_ECMA}, though this representation may affect the interpretation of the unobserved heterogeneity.

The (conditional on $\bX_i$) MTE at level $a$ is defined as 
\[
\MTE(a|\bx) = \Expectation[Y_i(1)-Y_i(0)|V_i=a,\bX_i=\bx].
\]
The MTE will be constant in $a$ if either (i) the individual treatment effect $Y_i(1)-Y_i(0)$ is constant, or (ii) there is no selection on unobservables, that is, the error terms of the outcome equation (\ref{eq: switching regression 1}) are unrelated to that of the selection equation (\ref{eq: switching regression 2}). The parameter $\MTE(a|\bx)$ is understood as the treatment effect for the subpopulation where an infinitesimal increase in the propensity score leads to a change in participation status. Note that for $a$ close to 1, the MTE measures the treatment effect in a subpopulation that is very \textit{unlikely} to be treated. Other treatment and policy effects can be recovered using the MTE.

Two assumptions are made to facilitate identification. First, the collection of instruments $\bZ_i$ is nondegenerate and independent of the error terms $(U_{1i},U_{0i},V_i)$ conditional on the covariates $\bX_i$. Second, $0<\Prob[T_i=1|\bX_i]<0$, so that conditional on the covariates, both treated and untreated individuals are observable in the population. It can then be shown that, for any limit point $a$ in the support of the propensity score, $\MTE(a|\bx)$ is
\[\MTE(a|\bx) = \frac{\partial}{\partial a}\Expectation[Y_i|P_i=a,\bX_i=\bx].\]
This representation shows that the MTE is identifiable, and could in principle be estimated by standard nonparametric techniques (once $P_i$ is estimated). In practice, however, nonparametric methods for estimating $\MTE(a|\bx)$ and functionals thereof are often avoided because of the curse of dimensionality, the negative impact of smoothing and tuning parameters, and efficiency considerations. A flexible parametric functional form can be used instead: $\Expectation[Y_i|P_i, \bX_i] = e(\bX_i,P_i,\btheta_0)$, where $e(\cdot)$ is a known function up to some finite dimensional parameter $\btheta_0$.

Therefore, the MTE estimator is often constructed as follows:
\begin{alignat*}{2}
\hat{\tau}_\mathtt{MTE}(a|\bx) &= \frac{\partial}{\partial a} e(\bx,a,\hat\btheta),
&&\qquad \hat\btheta = \argmin_{\btheta} \sum_{i=1}^n \Big(Y_i-e(\bX_i,\hat{P}_i,{\btheta})\Big)^2,\\
\hat{P}_i &= \bZ_i^\Trans\hat{\bbeta},
&&\qquad \hat{\bbeta}= \argmin_{\bbeta}\sum_{i=1}^n \Big( T_i - \bZ_i^\Trans\bbeta \Big)^2,
\end{alignat*}

Identification and estimation of the MTE, as well as other policy-relevant parameters based on it, require exogenous variation in the treatment equation (\ref{eq: switching regression 2}) induced by instrumental variables. In practice, researchers induce this variation by (i) employing many instruments, possibly generating them using power expansions and interactions, and (ii) including interactions with the ``raw'' or expanded instruments. Employing a flexible, high-dimensional specification for the probability of selection is also useful to mitigate misspecification errors. These observations have led researchers to employ many covariates/instruments in the probability of selection, that is, have a ``large'' $k$ relative to the sample size. In this paper, we show that flexibly modeling the probability of selection can lead to a first-order bias in the estimation of the MTE and related policy-relevant estimands, even when the outcome equation is modeled parametrically and low-dimensional. Furthermore, we provide automatic bias-correction and inference procedures based on resampling methods.

The following result characterizes the asymptotic properties of the estimated MTE.

\begin{coro}[Asymptotic Normality: MTE]\ \\
	Suppose the assumptions of Theorem \ref{Theorem: Asymptotic Representation} and the example-specific additional regularity conditions summarized in the Supplemental Appendix hold. Then, for $\hat{\btheta}$,
	\begin{align*}
	\mathbf{B}_{i} &= \frac{\partial^2 e(\bX_i,P_i,\btheta_0)}{\partial P_i\partial \btheta}\Big[ (1-P_i)\cdot\Expectation[T_iY_i(1)|\bZ_i] - P_i\cdot \Expectation[(1-T_i)Y_i(0)|\bZ_i] \Big]\pi_{ii} \\
	&\qquad +  \frac{1}{2}\sum_{j=1}^{n}\left[\frac{\partial^2 e(\bX_i,P_i,\btheta_0)}{\partial P_i\partial \btheta}\tau_{\mathtt{MTE}}(P_i|\bX_i)+\frac{1}{2}\frac{\partial e(\bX_i,P_i,\btheta_0)}{\partial \btheta}\frac{\partial \tau_{\mathtt{MTE}}(P_i|\bX_i)}{\partial P_i}\right]P_j(1-P_j)\pi_{ij}^2,\\
	\bPsi_i &= \frac{\partial e(\bX_i,P_i,\btheta_0)}{\partial \btheta}\Big(Y_i -e(\bX_i,P_i,\btheta_0) \Big) -\left(\sum_{j=1}^n \frac{\partial e(\bX_j,P_j,\btheta_0)}{\partial \btheta}\tau_{\mathtt{MTE}}(P_j|\bX_j)\pi_{ij}\right)(T_i-P_i),
	\end{align*}
	and $\bSigma_0$ is given in the Supplemental Appendix (Section SA-5.4). 
\end{coro}

The above result gives a precise characterization of the asymptotic possibly first-order bias and variance of $\hat{\btheta}$ via the results in Theorem \ref{Theorem: Asymptotic Representation}. To obtain the corresponding result for the estimated MTE, $\hat{\tau}_\mathtt{MTE}(a|\bx)$, the delta method is employed and an extra multiplicative factor $\partial^2 e(\bx,a,\btheta_0)/\partial a\partial \btheta^\Trans$ shows up. As a result, both the bias and variance for the estimated MTE will depend on the evaluation point $(\bx|a)$. We give the details in the Supplemental Appendix Section SA-5.4.

To understand the implications of the above corollary, we consider the bias terms. Note that the factor associated with $\pi_{ii}$ essentially captures treatment effect heterogeneity (in the outcome equation) and self-selection. To make it zero, one needs to assume there is no heterogeneous treatment effect and that the agents do not act on idiosyncratic characteristics that are unobservable to the analyst. For the second bias term associated with $\pi_{ij}^2$, note that it involves both the level of the MTE and its curvature. Hence the second bias is related not only to treatment effect heterogeneity captured through the shape of the MTE, but also to the magnitude of the treatment effect. Thus, aside from the off chance of these terms canceling each other, the many instruments bias will be zero only when there is neither heterogeneity nor self-selection, and the treatment effect is zero. Since these conditions are unlikely to hold in empirical work, even in randomized controlled trials, we expect the many instruments bias to have a direct implication in most practical cases. Therefore, conventional estimation and inference methods that do not account for the many instruments bias will be invalid, even in large samples, when many instruments are included in the estimation.

The results above take the conditional expectation function $e(\bX_i,P_i,\btheta_0)$ as low-dimensional, but in practice researchers may want to include many covariates also in the second estimation step. In Section SA-4.3 of the Supplemental Appendix, we study a generalization of Theorem \ref{Theorem: Asymptotic Representation} for the special case when $\Expectation[Y_i|P_i, \bX_i, \bW_i] = e(\bX_i,P_i,\btheta_0) + \bW_i^\Trans\bgamma_0$, where $\bW_i$ contains additional conditioning variables (possibly including $\bX_i$) and the nuisance parameter $\bgamma_0$ is potentially high-dimensional. If nonlinear least-squares is used to estimate the second-step as above, we find that additional terms now contribute to the many covariates bias due to the possibly high-dimensional estimation of $\bgamma_0$ in the second-step, but the same general results reported in this paper continue to hold. Specifically, the many covariates bias remains of order $k/\sqrt{n}$ and needs to be accounted for in order to conduct valid inference whenever $k/\sqrt{n}$ is not negligible.

\section{Numerical Evidence}\label{section: numerical}

We provide numerical evidence for the methods developed in this paper. First, we offer a Monte Carlo experiment constructed in the context of MTE estimation (Section \ref{section: MTE}), which highlights the role of the many covariates bias and showcases the role of jackknife bias correction and bootstrap approximation for estimation and inference. Second, also in the context of MTE estimation and inference, we offer an empirical illustration following the work of \cite*{Carneiro-Heckman-Vytlacil_2011_AER}. Section SA-8 of the Supplemental Appendix contains more results and further details omitted here to conserve space.

\subsection{Simulation Study}

We retain the notation and assumptions imposed in Section \ref{section: MTE}, and set the potential outcomes to $Y_i(0)=U_{0i}$ and $Y_i(1) = 0.5 + U_{1i}$. We assume there are many potential instruments $\bZ_i=\left[1,Z_{1,i},Z_{2,i},\dots,Z_{199,i}\right]$, with $Z_{\ell,i}\sim\mathrm{Uniform}[0,1]$ independent across $\ell=1,2,\dots,199$. The selection equation is assumed to take a very parsimonious form: $T_i=\Indicator\big[ 0.1+Z_{1,i}+Z_{2,i}+Z_{3,i}+Z_{4,i} \geq V_i\big]$. In this case Assumption \ref{Assumption: first-step} holds automatically without misspecification error, but in the Supplemental Appendix we explore other specifications of the propensity score where approximation errors are present. Finally, the error terms are distributed as $V_i|\bZ_i\thicksim\mathrm{Uniform}[0,1]$, $U_{0i}|\bZ_i,V_i\thicksim\mathrm{Uniform}[-1,1]$ and $U_{1i}|\bZ_i,V_i\thicksim\mathrm{Uniform}[-0.5, 1.5-2V_i]$. Because additional covariates $\bX_i$ do not feature in this data generating process, the treatment effect heterogeneity and self-selection are captured by the correlation between $U_{1i}$ and $V_i$.

It follows that $\Expectation[Y_i|P_i=a] = a - \frac{a^2}{2}$, and the MTE is $\MTE(a) = 1-a$. Given a random sample index by $i=1,2,\dots,n$, the second-step regression model is set to $\Expectation[Y_i|P_i]=\theta_1 + \theta_2\cdot P_{i} + \theta_3\cdot P^2_{i}$ and therefore the estimated MTE is $\hat{\tau}_{\mathtt{MTE}}(a) = \hat{\theta}_2 + 2a\cdot \hat{\theta}_3$ with $(\hat{\theta}_1,\hat{\theta}_2,\hat{\theta}_3)'$ denoting the least-squares estimators of $({\theta}_1,{\theta}_2,{\theta}_3)'$. We consider the quantity $\sqrt{n}\left( \hat\tau_{\mathtt{MTE}}(a) - \tau_{\mathtt{MTE}}(a) \right)$ at $a=0.5$, with and without bias correction, for two sample sizes $n=1,000$ and $n=2,000$, and across $2,000$ simulation repetitions. To estimate the propensity score, we regress $T_i$ on a constant term and $\{Z_{\ell,i}\}$ for $1\leq \ell\leq k-1$, where the number of covariates $k$ ranges from $5$ to $200$. Note that $k=5$ corresponds to the most parsimonious model which is correctly specified.  

For inference, we consider two approaches. In the conventional approach, the many instruments bias is ignored, and hypothesis testing is based on normal approximation to the t-statistic, where the standard error comes from the simulated sampling variability of the estimator (i.e. the oracle standard error, which is infeasible). That is, this benchmark approach considers the infeasible statistic $(\hat{\tau}_{\mathtt{MTE}}-\tau_{\mathtt{MTE}})/\sqrt{\Var[\hat{\tau}_{\mathtt{MTE}}]}$, with $\Var[\hat{\tau}_{\mathtt{MTE}}]$ denoting the simulation variance of $\hat{\tau}_{\mathtt{MTE}}$, and employs standard normal quantiles. The other approach, which follows the results in this paper, utilizes both the jackknife and the bootstrap: the feasible statistic $(\hat{\tau}_{\mathtt{MTE}}-\hat{\mathscr{B}}-\tau_{\mathtt{MTE}})/\sqrt{\hat{\mathscr{V}}}$ is constructed as in Section \ref{section: jackknife} and inference is conducted using the bootstrap approximation as in Section \ref{section: bootstrap}. 

The results are collected in Table \ref{table: MTE-2-BT-JK}. The bias is small with small $k$, as the most parsimonious model is correctly specified. With more instruments added to the propensity score estimation, the many instruments bias quickly emerges, and without bias correction, it leads to severe empirical undercoverage (conventional $95\%$ confidence is used). Interestingly, the finite sample variance shrinks at the same time. Therefore for this particular DGP, incorporating many instruments not only leads to biased estimates, but also gives the illusion that the parameter is estimated precisely. With jackknife bias correction, there is much less empirical size distortion, and the empirical coverage rate remains well-controlled even with $200$ instruments used in the first step. Moreover, the jackknife bias correction also (partially) restores the true variability of the estimator.  

Although the focus here is on inference and, in particular, empirical coverage of associated testing procedures, it is also important to know how the bias correction will affect the Standard Deviation (sd) and the Mean Squared Error (MSE) of the point estimators. Recall that the model is correctly specified with $5$ instruments, hence it should not be surprising that incorporating bias correction there increases the variability of the estimator and the MSE -- although the impact is very small. As more instruments are included, however, the MSE increases rapidly without bias correction, while the MSE of the bias corrected estimator remains relatively stable. In particular, this finding is driven by a sharp reduction in bias that more than compensates the increase in variability of the estimator. A larger variance of the bias-corrected estimator is expected, as additional sampling variability is introduced by the bias correction. All in all, the bias-corrected estimator seems to be appealing not only for inference, but also for point estimation because it performs better in terms of MSE when the number of instruments is moderate or large. 

In the Supplemental Appendix, we report results from two other data generating processes. In particular, we consider cases when (i) the true propensity score is non-linear and fundamentally misspecified; and (ii) the true propensity score is non-linear and low-dimensional, and one employs basis expansion to approximate the true propensity score. The exact magnitude of the bias changes in different settings, but the same pattern emerges: as the number of included instruments/basis elements increases, the asymptotic distribution is no longer centered at the true parameter due to the bias uncovered in this paper (Theorem \ref{Theorem: Asymptotic Representation}). Moreover, the jackknife continues to provide excellent bias correction (Theorem \ref{Theorem: Jackknife}), and the bootstrap performs very well in approximating the finite sample distribution (Theorem \ref{Theorem: bootstrap+jackknife}). 

\subsection{Empirical Illustration}

In this section we consider estimating the marginal returns to college education following the work of \citet[][CHV hereafter]{Carneiro-Heckman-Vytlacil_2011_AER} with MTE methods, employing the notation and assumptions imposed in Section \ref{section: MTE}. The data consists of a subsample of white males from the 1979 National Longitudinal Survey of Youth (NLSY79), and the sample size is $n=1,747$. The outcome variable, $Y_i$, is the log wage in 1991, and the sample is split according to the treatment variable $T_i=0$ (high school dropouts and high school graduates), and $T_i=1$ (with some college education or college graduates). The dataset includes covariates on individual and family background information, and four ``raw'' instrumental variables: presence of four-year college, average tuition, local unemployment and wage rate, measured at age 17 of the survey participants.

We normalize the estimates by the difference of average education level between the two groups, so that the estimates are interpreted as the return to per year of college education. We make the same assumption as in CHV that the error terms are jointly independent of the covariates and the instruments. Then, $\tau_{\mathtt{MTE}}(a|\bx) = \partial \Expectation[Y_i|P_i=a,\bX_i=\bx]/\partial a$ with
\[\Expectation[Y_i|P_i=a,\bX_i=\bx] =  \bx^\Trans \bgamma_{0} + a\cdot \bx^\Trans \bdelta_{0} + \boldsymbol{\phi}(a)^\Trans \btheta_{0},\]
where $P_i=\Prob[T_i=1|\bZ_i]$ is the propensity score, and $\boldsymbol{\phi}$ is some fixed transformation. The covariates $\bX_i$ include (i) linear and square terms of corrected AFQT score, education of mom, number of siblings, permanent average local unemployment rate and wage rate at age 17; (ii) indicator of urban residency at age 14; (iii) cohort dummy variables; and (iv) average local unemployment rate and wage rate in 1991, and linear and square terms of work experience in 1991. For the selection equation, the instruments $\bZ_i$ include (i), (ii) and (iii) described earlier, as well as (v) the four raw instruments as well as their interactions with corrected AFQT score, education of mom and number of siblings. To make the functional form of the propensity score flexible, we also include interactions among the variables described in (i), and interactions between the cohort dummies and corrected AFQT score, education of mom and number of siblings. To conserve space, we leave summary statistics to the Supplemental Appendix. 

We are employing the same covariates, instruments, and modeling assumptions as in CHV, but our estimation strategy is different than theirs. For the first step, the selection equation (propensity score) is estimated using a linear probability model with $k=66$ as more interaction terms are included (which implies $k/\sqrt{n}=1.58$), while CHV employ a Logit model with $k=35$. Thus, our estimation approach reflects Assumption \ref{Assumption: first-step} in the sense that we assume away misspecification errors from using a flexible (high-dimensional) linear probability model, while CHV assume away misspecification errors from using a lower dimensional Logit model. For the second step, while the specification of $\Expectation[Y_i|P_i=a,\bX_i=\bx]$ coincides, we estimate the partially linear model (that is, the $\bphi(a)$ component) using a flexible polynomial in $P_i$ while CHV employ a kernel local polynomial approach with a bandwidth of about $0.30$ over the support $[0,1]$. To be specific, we implement the second step estimation by using least-squares regression with a fourth-order polynomial of the estimated propensity score $\bphi(\hat{P}_i) = [\hat{P}_i, \hat{P}_i^2,\hat{P}_i^3,\hat{P}_i^4]^\Trans$. Here the dimension of $\bX_i$ is $23$, so the second step model can be regarded as either ``flexible'' parametric or high-dimensional. In the latter case, the results reported in Section SA-4.3 of the Supplemental Appendix can be used, together with standard results from high-dimensional linear regression \citep[see][and references therein]{Cattaneo-Jansson-Newey_2018_ET,Cattaneo-Jansson-Newey_2018_JASA}, to show that a many covariate bias continues to be present in this setting, thereby justifying the usefulness of our fully automatic bias-correction and bootstrap-based methods. Finally, also in the Supplemental Appendix, we give results for other specifications of the selection and outcome equations.

\begin{figure}[!htp]
\centering
\includegraphics[width=10cm]{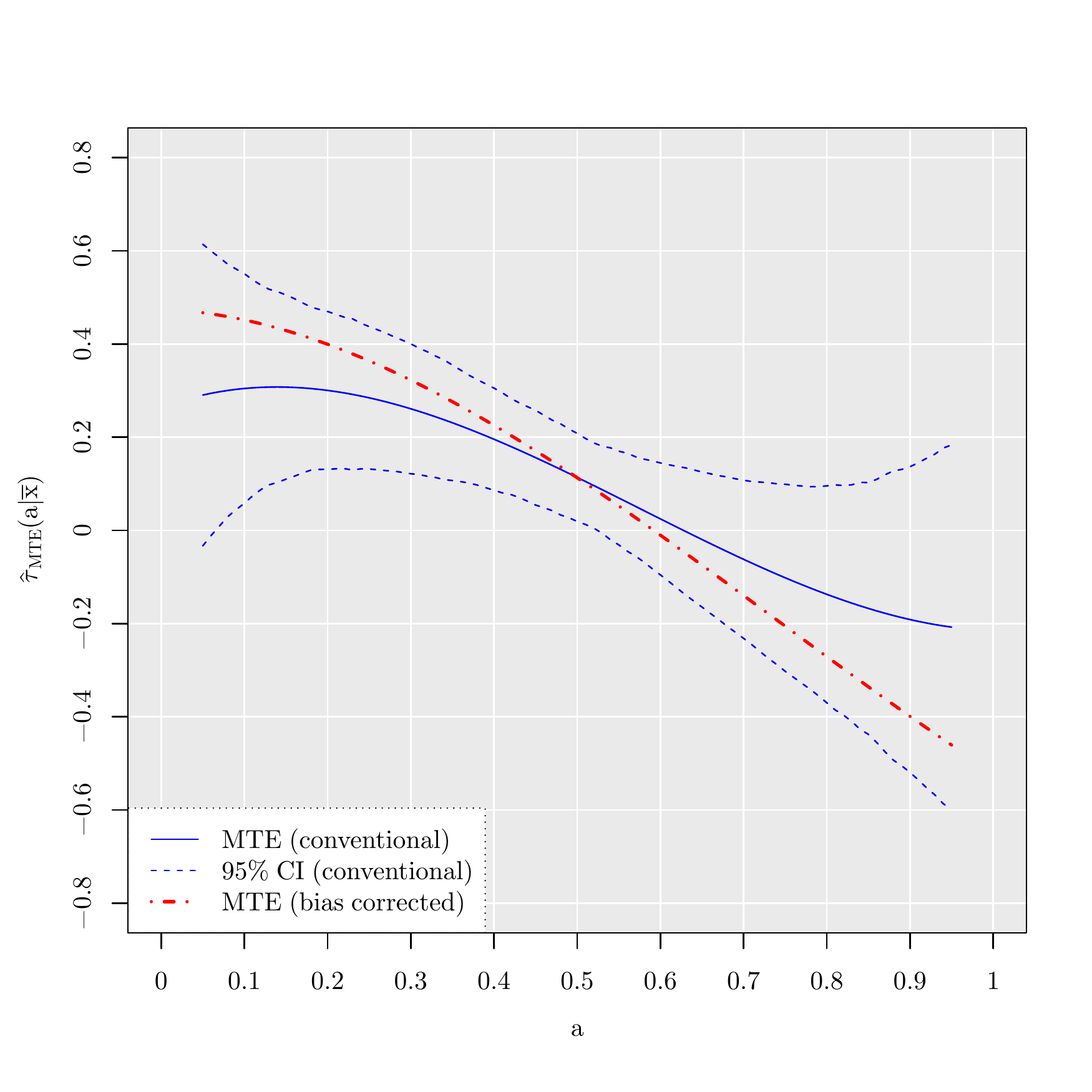}
\vskip -1cm
\includegraphics[width=10cm]{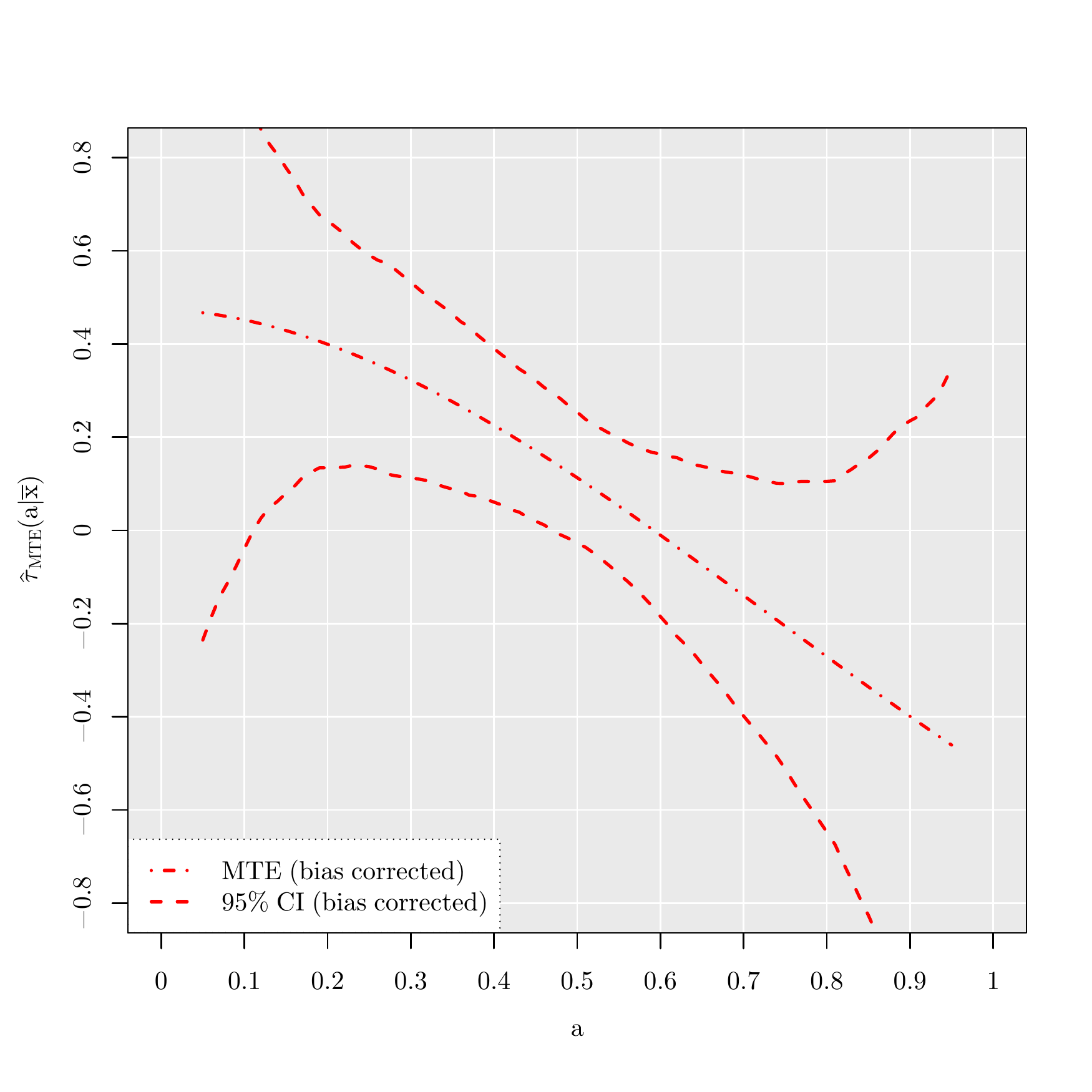}
\caption{Marginal Treatment Effects}\label{Fig: empirical}
{\footnotesize\begin{flushleft}
The marginal treatment effect, $\hat{\tau}_{\mathtt{MTE}}(a|\bar{\bX})$, is evaluated at mean value of the covariates. Bootstrap is used to construct the confidence interval, with 500 repetitions. \\
\textbf{Upper panel.} Estimated MTE without bias correction (solid blue line), together with 95\% confidence interval (dashed blue line). Also included is the bias-corrected MTE (dashed red line).\\
\textbf{Lower panel.} Bias-corrected MTE, together with 95\% confidence interval, taking into account the effect of bias correction.
\end{flushleft}}
\end{figure}

We summarize the empirical findings in Figure \ref{Fig: empirical}, where we plot the estimated MTE evaluate at the sample average of $\bX_i$. In the upper panel of this figure, we plot the estimated MTE together with $95\%$ confidence intervals (solid and dashed blue line), using conventional two-step estimation methods (i.e., without bias correction and employing the standard normal approximation). These empirical results are quite similar to those presented by CHV, both graphically and numerically. In particular, for individuals who are very likely to enroll in college, the per year return can be as high as $30\%$, while the return to college can also be as low as $-20\%$ for people who are very unlikely to enroll. Integrating the estimated MTE gives an estimator of the average treatment effect, which is roughly $9\%$.

The upper panel of Figure \ref{Fig: empirical}, also depicts the bias-corrected MTE estimator (dashed red line). The average treatment effect corresponding to the bias-corrected MTE is $8\%$, quite close to the previous estimate. On the other hand, the bias-corrected MTE curve has much steeper slope, implying a wider range of heterogeneity for returns to college education. This bias-corrected MTE curve lies close to the boundary of the confidence intervals constructed using the conventional two-step method, hinting at the possibility of a many instruments/covariate bias in the conventional estimate (blue line).

The lower panel of Figure \ref{Fig: empirical} plots the bias corrected MTE estimator, together with the confidence intervals constructed using our proposed bootstrap-based method, which takes into account the extra variability introduced by bias correction. Not surprisingly, the new confidence intervals are wider than the conventional ones.

\section{Conclusion}\label{section: conclusion}

We studied the distributional properties of two-step estimators, and functionals thereof, when possibly many covariates are used to fit the first-step estimate (e.g., a propensity score, generated regressors or control functions). We show that overfitting in the first step estimation leads to a first-order bias in the distributional approximation of the two-step estimator. As a consequence, the limiting distribution is no longer centered at zero and usual inference procedures become invalid, possibly exhibiting severe empirical size distortions in finite samples. We considered a few extensions of our basic framework and illustrated our generic results with several applications in treatment effect, program evaluation and other applied microeconomic settings. In particular, we presented new results for estimation and inference in the context of production function and marginal treatment effects estimation. The latter application, along with the one on local average response functions discussed in the Supplemental Appendix, give new results in the context of IV models with treatment effect heterogeneity and many instruments, previously unavailable in the literature.

As a remedy for the many covariates bias we uncover, we develop bias correction methods using the jackknife. Importantly, this approach is data-driven and fully automatic, and does not require additional resampling beyond what would be needed to compute the jackknife standard error, which we show is also consistent in our setting even when many covariates are used. Therefore, implementation is straightforward and is available in any statistical computing software. Furthermore, to improve finite sample inference after bias-correction, we also establish validity of an appropriately modified bootstrap for the jackknife-based bias-corrected Studentized statistic. We demonstrate the performance of our estimation and inference procedures in a comprehensive simulation study and an empirical illustration.

From a more general perspective, our main results give one additional contribution. They shed new light on the ultra-high-dimensional literature: one important implication is that typical sparsity assumptions imposed in that literature cannot be dropped in the context of non-linear models, since otherwise the effective number of \textit{included} covariates will remain large after model selection, which in turn will lead to a non-vanishing first-order bias in the distributional approximation for the second-step estimator. It would be interesting to explore whether resampling methods are able to successfully remove this many \textit{selected} or \textit{included} covariates bias in ultra-high-dimensional settings, where model selection techniques are also used as a first-step estimation device.


\begin{onehalfspace}
\addcontentsline{toc}{section}{References}
\bibliographystyle{econometrica}
\bibliography{Cattaneo-Jansson-Ma_2018_ManyCovs}
\end{onehalfspace}
\clearpage

\clearpage

\begin{table}[p]
\renewcommand{\arraystretch}{1.3}
\centering
\caption{Simulation: Marginal Treatment Effects}\label{table: MTE-2-BT-JK}
\subfloat[$n=1000$]{\resizebox{0.9\columnwidth}{!}{\input{summary1000.txt}}}\\ 
\subfloat[$n=2000$]{\resizebox{0.9\columnwidth}{!}{\input{summary2000.txt}}}
{\footnotesize\begin{flushleft}
\textbf{Notes}. The marginal treatment effect is evaluated at $a=0.5$. Panel (a) and (b) correspond to sample size $n=1000$ and $2000$, respectively. Statistics are centered at the true value. $k=5$ is the correctly specified model. 

\textbf{(i)} $k$: number of instruments used for propensity score estimation; \\ 
\textbf{(ii)} bias: empirical bias (scaled by $\sqrt{n}$); \\
\textbf{(iii)} sd: empirical standard deviation (scaled by $\sqrt{n}$); \\ 
\textbf{(iv)} $\sqrt{\text{mse}}$: empirical root-MSE (scaled by $\sqrt{n}$); \\
\textbf{(v)} coverage: empirical coverage of a $95\%$ confidence interval. Without bias correction, it is based on normal approximation and simulated sampling variability of the estimator (i.e. the oracle standard error). With bias correction, the test is based on the percentile-t method, where the bias-corrected and Studentized statistic is bootstrapped 500 times (Rademacher weights);\\
\textbf{(vi)} length: the average confidence interval length (scaled by $\sqrt{n}$).
\end{flushleft}}
\end{table}
\clearpage

\clearpage

\end{document}

%% file: summary1000.txt
\begin{tabular}{lrrcccccccccccc}
\hline\hline
\multicolumn{1}{l}{\bfseries }&\multicolumn{2}{c}{\bfseries }&\multicolumn{1}{c}{\bfseries }&\multicolumn{5}{c}{\bfseries Conventional}&\multicolumn{1}{c}{\bfseries }&\multicolumn{5}{c}{\bfseries Bias-Corrected}\tabularnewline
\cline{5-9} \cline{11-15}
\multicolumn{1}{l}{}&\multicolumn{1}{c}{$k/n$}&\multicolumn{1}{c}{$k/\sqrt{n}$}&\multicolumn{1}{c}{}&\multicolumn{1}{c}{bias}&\multicolumn{1}{c}{sd}&\multicolumn{1}{c}{$\sqrt{\text{mse}}$}&\multicolumn{1}{c}{coverage}&\multicolumn{1}{c}{length}&\multicolumn{1}{c}{}&\multicolumn{1}{c}{bias}&\multicolumn{1}{c}{sd}&\multicolumn{1}{c}{$\sqrt{\text{mse}}$}&\multicolumn{1}{c}{coverage}&\multicolumn{1}{c}{length}\tabularnewline
\hline
{\bfseries $k$}&&&&&&&&&&&&&&\tabularnewline
~~5&$0.00$&$0.16$&&$0.14$&$4.72$&$4.73$&$0.95$&$18.51$&&$-0.21$&$4.93$&$4.93$&$0.93$&$18.28$\tabularnewline
~~20&$0.02$&$0.63$&&$1.73$&$4.11$&$4.46$&$0.93$&$16.11$&&$ 0.18$&$5.26$&$5.27$&$0.94$&$19.81$\tabularnewline
~~40&$0.04$&$1.26$&&$3.08$&$3.54$&$4.69$&$0.86$&$13.88$&&$ 1.03$&$5.11$&$5.22$&$0.94$&$19.67$\tabularnewline
~~60&$0.06$&$1.90$&&$3.96$&$3.22$&$5.11$&$0.77$&$12.63$&&$ 1.75$&$5.02$&$5.32$&$0.93$&$19.27$\tabularnewline
~~80&$0.08$&$2.53$&&$4.61$&$3.00$&$5.50$&$0.66$&$11.76$&&$ 2.28$&$4.91$&$5.41$&$0.92$&$18.67$\tabularnewline
~~100&$0.10$&$3.16$&&$5.10$&$2.83$&$5.83$&$0.56$&$11.08$&&$ 2.65$&$4.78$&$5.46$&$0.90$&$18.28$\tabularnewline
~~120&$0.12$&$3.79$&&$5.55$&$2.67$&$6.16$&$0.46$&$10.48$&&$ 2.96$&$4.66$&$5.51$&$0.89$&$17.80$\tabularnewline
~~140&$0.14$&$4.43$&&$5.97$&$2.54$&$6.49$&$0.35$&$ 9.98$&&$ 3.24$&$4.57$&$5.60$&$0.87$&$17.46$\tabularnewline
~~160&$0.16$&$5.06$&&$6.35$&$2.45$&$6.81$&$0.26$&$ 9.59$&&$ 3.46$&$4.43$&$5.62$&$0.86$&$17.15$\tabularnewline
~~180&$0.18$&$5.69$&&$6.69$&$2.33$&$7.09$&$0.18$&$ 9.13$&&$ 3.58$&$4.35$&$5.63$&$0.86$&$16.97$\tabularnewline
~~200&$0.20$&$6.32$&&$7.03$&$2.23$&$7.38$&$0.12$&$ 8.75$&&$ 3.81$&$4.22$&$5.69$&$0.84$&$16.75$\tabularnewline
\hline
\end{tabular}

%% file: summary2000.txt
\begin{tabular}{lrrcccccccccccc}
\hline\hline
\multicolumn{1}{l}{\bfseries }&\multicolumn{2}{c}{\bfseries }&\multicolumn{1}{c}{\bfseries }&\multicolumn{5}{c}{\bfseries Conventional}&\multicolumn{1}{c}{\bfseries }&\multicolumn{5}{c}{\bfseries Bias-Corrected}\tabularnewline
\cline{5-9} \cline{11-15}
\multicolumn{1}{l}{}&\multicolumn{1}{c}{$k/n$}&\multicolumn{1}{c}{$k/\sqrt{n}$}&\multicolumn{1}{c}{}&\multicolumn{1}{c}{bias}&\multicolumn{1}{c}{sd}&\multicolumn{1}{c}{$\sqrt{\text{mse}}$}&\multicolumn{1}{c}{coverage}&\multicolumn{1}{c}{length}&\multicolumn{1}{c}{}&\multicolumn{1}{c}{bias}&\multicolumn{1}{c}{sd}&\multicolumn{1}{c}{$\sqrt{\text{mse}}$}&\multicolumn{1}{c}{coverage}&\multicolumn{1}{c}{length}\tabularnewline
\hline
{\bfseries $k$}&&&&&&&&&&&&&&\tabularnewline
~~5&$0.00$&$0.11$&&$0.13$&$4.85$&$4.85$&$0.95$&$19.00$&&$-0.12$&$4.95$&$4.95$&$0.93$&$18.21$\tabularnewline
~~20&$0.01$&$0.45$&&$1.42$&$4.47$&$4.69$&$0.94$&$17.51$&&$ 0.06$&$5.16$&$5.16$&$0.94$&$19.31$\tabularnewline
~~40&$0.02$&$0.89$&&$2.73$&$4.17$&$4.99$&$0.90$&$16.36$&&$ 0.54$&$5.35$&$5.38$&$0.94$&$19.72$\tabularnewline
~~60&$0.03$&$1.34$&&$3.78$&$3.95$&$5.47$&$0.84$&$15.47$&&$ 1.18$&$5.44$&$5.57$&$0.93$&$19.75$\tabularnewline
~~80&$0.04$&$1.79$&&$4.62$&$3.74$&$5.95$&$0.76$&$14.67$&&$ 1.82$&$5.43$&$5.73$&$0.91$&$19.59$\tabularnewline
~~100&$0.05$&$2.24$&&$5.27$&$3.55$&$6.35$&$0.68$&$13.91$&&$ 2.33$&$5.37$&$5.86$&$0.90$&$19.31$\tabularnewline
~~120&$0.06$&$2.68$&&$5.77$&$3.37$&$6.68$&$0.59$&$13.22$&&$ 2.74$&$5.27$&$5.94$&$0.90$&$19.04$\tabularnewline
~~140&$0.07$&$3.13$&&$6.27$&$3.20$&$7.03$&$0.49$&$12.53$&&$ 3.21$&$5.11$&$6.04$&$0.88$&$18.85$\tabularnewline
~~160&$0.08$&$3.58$&&$6.67$&$3.07$&$7.35$&$0.41$&$12.03$&&$ 3.53$&$5.05$&$6.16$&$0.87$&$18.66$\tabularnewline
~~180&$0.09$&$4.02$&&$7.07$&$2.95$&$7.65$&$0.32$&$11.54$&&$ 3.87$&$4.95$&$6.28$&$0.85$&$18.40$\tabularnewline
~~200&$0.10$&$4.47$&&$7.42$&$2.83$&$7.94$&$0.26$&$11.11$&&$ 4.13$&$4.84$&$6.36$&$0.85$&$18.22$\tabularnewline
\hline
\end{tabular}